\documentclass[letterpaper,10pt]{article}
\usepackage[letterpaper,textwidth=5.5in,textheight=9in,centering,footskip=0.5in]{geometry}
\usepackage{amsmath}
\usepackage{iftex}
\ifPDFTeX
  \usepackage[T1]{fontenc}
  \usepackage[utf8]{inputenc}
  \usepackage{tgtermes}
  \usepackage{pifont}
  \usepackage{newunicodechar}
  \newunicodechar{✓}{\ding{51}}
  \newunicodechar{✗}{\ding{55}}
  \newunicodechar{‐}{-}
\else
  \usepackage{fontspec}
  \usepackage{unicode-math}
  \newfontface\symbolfont{DejaVuSans.ttf}[Scale=MatchLowercase]
  \usepackage{newunicodechar}
  \newunicodechar{✓}{{\symbolfont ✓}}
  \newunicodechar{✗}{{\symbolfont ✗}}
\fi
\newunicodechar{η}{\ensuremath{\eta}}
\newunicodechar{γ}{\ensuremath{\gamma}}
\newunicodechar{λ}{\ensuremath{\lambda}}
\newunicodechar{Δ}{\ensuremath{\Delta}}
\newunicodechar{ω}{\ensuremath{\omega}}
\newunicodechar{≥}{\ensuremath{\geq}}
\newunicodechar{≤}{\ensuremath{\leq}}
\newunicodechar{→}{\ensuremath{\rightarrow}}
\newunicodechar{×}{\ensuremath{\times}}
\newunicodechar{−}{\ensuremath{-}}
\newunicodechar{≈}{\ensuremath{\approx}}
\newunicodechar{±}{\ensuremath{\pm}}
\newunicodechar{·}{\ensuremath{\cdot}}
\newunicodechar{µ}{\ensuremath{\mu}}
\newunicodechar{μ}{\ensuremath{\mu}}
\newunicodechar{α}{\ensuremath{\alpha}}
\newunicodechar{β}{\ensuremath{\beta}}
\newunicodechar{σ}{\ensuremath{\sigma}}
\newunicodechar{θ}{\ensuremath{\theta}}
\newunicodechar{​}{}
\newunicodechar{⁻}{\textsuperscript{-}}
\newunicodechar{⁴}{\textsuperscript{4}}
\newunicodechar{⁵}{\textsuperscript{5}}
\newunicodechar{⁶}{\textsuperscript{6}}
\newunicodechar{⁷}{\textsuperscript{7}}
\newunicodechar{⁸}{\textsuperscript{8}}
\newunicodechar{⁹}{\textsuperscript{9}}
\newunicodechar{⁰}{\textsuperscript{0}}
\usepackage[english]{babel}
\usepackage{microtype}
\usepackage{graphicx}
\usepackage{tikz}
\usepackage{rotating}
\usepackage{pdfpages}
\ifPDFTeX
  \makeatletter
  \newcommand{\landscapepage}{\protected@write\@auxout{}{\string\landscape@page{\thepage}}}
  \newcommand{\landscape@page}[1]{\expandafter\gdef\csname landscape@#1\endcsname{}}
  \AddToHook{shipout/before}{\ifcsname landscape@\thepage\endcsname
    \global\pdfpageattr{/Rotate 90}\else\global\pdfpageattr{}\fi}
  \makeatother
\else
  \newcommand{\landscapepage}{}
\fi
\usepackage[hidelinks]{hyperref}
\usepackage{titlesec}
\titleformat{\section}{\bfseries\large}{\thesection}{1em}{}
\titleformat{\subsection}{\bfseries\normalsize}{\thesubsection}{1em}{}
\titleformat{\subsubsection}{\bfseries\normalsize}{\thesubsubsection}{1em}{}
\titlespacing*{\section}{0pt}{2ex plus 0.5ex minus 0.2ex}{1.5ex plus 0.3ex}
\titlespacing*{\subsection}{0pt}{1.8ex plus 0.5ex minus 0.2ex}{0.8ex plus 0.2ex}
\titlespacing*{\subsubsection}{0pt}{1.5ex plus 0.5ex minus 0.2ex}{0.5ex plus 0.2ex}
\usepackage{caption}
\newcommand{\papertitle}[1]{\begin{center}{\fontsize{17}{20}\selectfont\bfseries #1\par}\end{center}\vspace{0.1in}}
\newcommand{\authors}[1]{\begin{center}\setlength{\parskip}{0pt}{\normalsize #1\par}\end{center}\vspace{-0.5ex}}
\newcommand{\affiliations}[1]{\begin{list}{}{\small\setlength{\leftmargin}{0.9em}\setlength{\labelwidth}{0.7em}%
  \setlength{\labelsep}{0.2em}\setlength{\itemsep}{1.5pt}\setlength{\parsep}{0pt}\setlength{\topsep}{6pt}%
  \raggedright}#1\end{list}}
\newcommand{\abstractblock}[1]{\vspace{0.15in}\begin{center}\textbf{Abstract}\end{center}\vspace{-0.5ex}%
  \begin{center}\begin{minipage}{\dimexpr\textwidth-1in\relax}\small #1\end{minipage}\end{center}\vspace{0.1in}}
\newcommand{\referencestyle}{\small\setlength{\parindent}{0pt}\setlength{\parskip}{1.5pt}}

\usepackage{flafter}

\begin{document}
\papertitle{APOD: reasoning-guided agentic population ordinary differential equation discovery for pharmacological digital twins}
\authors{Romain~Ferrara\textsuperscript{1,2,†}, Martin~Soucail\textsuperscript{1,†}, Victor~Gertner\textsuperscript{3}, Adil~Moussali\textsuperscript{2},\\Joris~Cocquebert\textsuperscript{2}, Sandrine~Oziel-Taieb\textsuperscript{4}, Julien~Nicolas\textsuperscript{5},\\Florence~Gattacceca\textsuperscript{1}, Mihaela~van~der~Schaar\textsuperscript{6}, Sébastien~Benzekry\textsuperscript{1}}
\affiliations{\item[\textsuperscript{1}] COMPutational pharmacology and clinical Oncology (COMPO), Inria Center at University Côte d'Azur, Cancer Research Center of Marseille, Marseille,~France
\item[\textsuperscript{2}] Department of Nuclear Medicine, Institut Paoli-Calmettes, ENETS Center of Excellence, IPC NET Center, Marseille,~France
\item[\textsuperscript{3}] Centre for Computational Biology (CBIO), Mines Paris, PSL University, Paris,~France
\item[\textsuperscript{4}] Department of Medical Oncology, Institut Paoli-Calmettes, ENETS Center of Excellence, IPC NET Center, Marseille,~France
\item[\textsuperscript{5}] Université Paris-Saclay, CNRS, Institut Galien Paris-Saclay, Orsay,~France
\item[\textsuperscript{6}] Department of Applied Mathematics and Theoretical Physics, University of Cambridge, Cambridge, United~Kingdom
\item[\textsuperscript{†}] These authors contributed equally to this~work.}
\abstractblock{Establishing ordinary differential equations (ODEs) describing population data is a fundamental part of mathematical modeling in pharmacology, crucial to developing digital twins. However, doing so from sparse, noisy data is a slow, expert-driven task. Existing automated methods either search a restricted model space or ignore population inter-individual variability. Here we introduce APOD (Agentic Population ODE Discovery), a language-model agent that iteratively reasons over biological knowledge and fit diagnostics in an open-ended search space to discover a population digital twin (PDT), i.e., a shared ODE system with between-subject variability. On synthetic pharmacokinetic and tumor-dynamics benchmarks, APOD recovered ground-truth structures in 94-100\% of runs, 12-fold faster in median than an established library-based search. On real cohorts it converged to valid structures, and proposed a PDT of radioligand-therapy-induced platelet dynamics that predicts thrombocytopenia from first-cycle data and simulates alternative dosing schedules that lower the predicted risk of toxicity.}
\section{Introduction}

Digital twins (DTs) are computational representations of a system that are continuously informed by data and used to predict its behavior and guide intervention\textsuperscript{1}. Ordinary differential equations (ODEs) are a natural language for such models: by expressing the rate of change of each quantity as a function of the others, they encode an explicit, causal structure rather than a statistical association. When fitted to data (and continuously updated), they allow forward simulations under counterfactual interventions. A pharmacological DT is a patient-specific virtual model that replicates an individual's pharmaco-physiology, is updated as new measurements arrive, predicts clinically relevant outcomes, and informs treatment decisions\textsuperscript{2--6}. They have two major applications. At the individual level, pharmacological DTs are used for dosing adaptation under therapeutic drug monitoring (model-informed precision dosing)\textsuperscript{7}. At the population level, they are used to determine the appropriate dosing regimen in clinical trials (model-informed drug discovery and development)\textsuperscript{8} and are now being investigated as synthetic control arms. As expressed by previous authors\textsuperscript{2}, one of the biggest challenges for medical DTs is accounting for the biological heterogeneity of patients.

Building ODE models has traditionally been a manual, time-consuming process: a modeler posits candidate equations from domain knowledge, and confronts them with data. Scientific machine learning methods, e.g. symbolic\textsuperscript{9,10} or sparse\textsuperscript{11} regression, have been proposed to automate the recovery of governing equations directly from observations\textsuperscript{12}. However, two features limit their reach in biomedicine. First, they operate on one subject at a time, whereas biomedical datasets are often composed of sparse, noisy and partial observations of a mechanism shared across multiple subjects in a population. Second, the search is framed as an optimization problem, driven by a single numeric objective that captures adequacy to the data and parsimony, but not biological plausibility.

A different strategy has recently emerged. Large Language Models (LLMs) have provided significant advances in artificial intelligence, with both advanced abilities for scientific discovery\textsuperscript{13} and mathematical and programming tasks\textsuperscript{14}. For ODEs, LLMs have been shown to be able to discover models autonomously, proposing equations from a natural-language prompt and iterating over fit diagnostics\textsuperscript{15,16}, including in the pharmacological field\textsuperscript{17}. Existing agents, however, remain far from clinical use. As above, they consider individual trajectories with no population-level inference; they still operate on a single goodness-of-fit metric; and they return an ODE of pre-defined dimension, precluding discovery of models with a different number of latent components. In addition, they have been tested only on data-rich synthetic systems\textsuperscript{15,16}, and rarely on sparse, heterogeneous cohorts where mechanistic discovery would carry clinical value\textsuperscript{17}.

The population approach in pharmacology was built to resolve the question of sparse sampling in individual subjects\textsuperscript{18}. In this framework -- nonlinear mixed-effects (NLME) modelling -- all subjects are assumed to share one mechanistic structure (the ODE), while each subject's parameters are drawn from a population-level distribution\textsuperscript{19}. An ODE model endowed with such a population structure -- which we term a population digital twin (PDT) -- supports the two operations a twin needs: subject-level prediction, in which the fitted population serves as a Bayesian prior that is personalized from sparse measurements to forecast an endpoint of interest, such as the probability of treatment toxicity, and simulation of counterfactual scenarios, such as an alternative dosing regimen.

Automated model discovery within the NLME framework is restricted to pharmacokinetics (PK), and follows the concepts of symbolic regression. Symbolic regression being an NP-hard problem\textsuperscript{20}, the established tools -- pyDarwin\textsuperscript{21--23}, Pharmpy's modelsearch\textsuperscript{24} and mlxModelFinder\textsuperscript{25} -- search a predefined library of PK building blocks. Thus, they cannot propose mechanisms outside it and a single search can take a few hours to several days per drug\textsuperscript{23}. In addition, the computational cost is substantial. Neural-ODE approaches relax the library constraint by fitting the governing function in a high-dimensional latent space\textsuperscript{26--29}, but need dense per-subject sampling and large cohorts to train and have so far remained case-specific. No existing method combines the two capabilities a population digital twin actually requires -- an open structural space, and population-level inference from sparse data (Table~1).

\begin{figure}[bp]
\centering
\caption*{\textbf{Table 1: Automated model-discovery approaches against the five properties a population digital twin requires.} Interpretable parameters: the parameters are pharmacological quantities, such as a clearance or a carrying capacity. Population-level: subjects share a structural model and a between-subject variability model. Small cohorts: usable on a few dozen sparsely sampled subjects. Fast method: the search completes in minutes rather than hours. Open structural space: candidates are not confined to a predefined library. ✓ yes, ✗ no, -- not applicable. A property that holds only in a restricted sense is scored yes. RF: random forest. XGBoost: extreme gradient boosting. RNN: recurrent neural network. ODE: ordinary differential equation. ML: machine learning. LLM: large language model. AMD: automatic model development.}
\includegraphics[width=\textwidth,trim=0 0 0 18bp,clip]{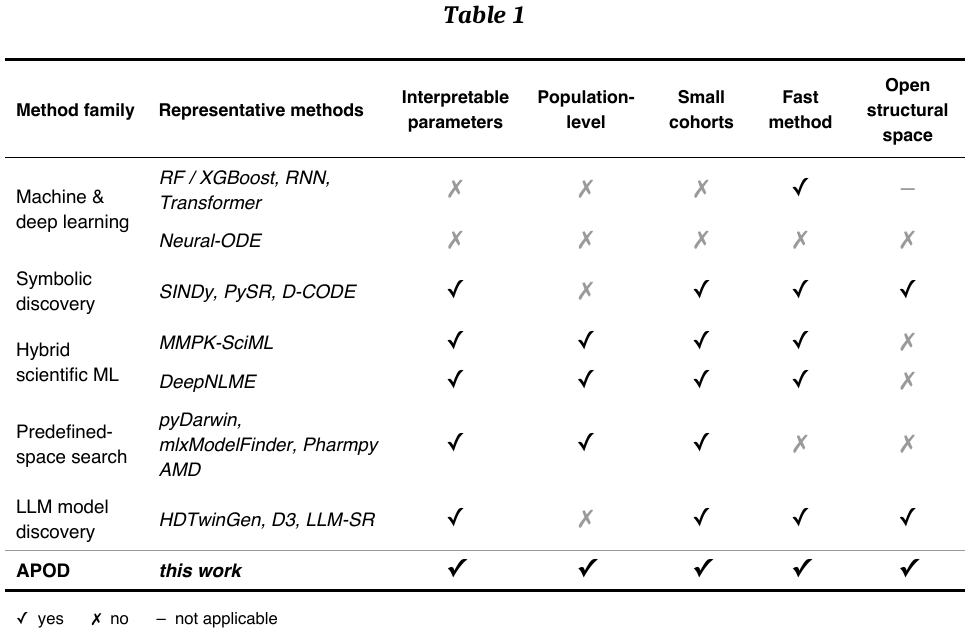}
\end{figure}

Here, we introduce APOD (Agentic Population ODE Discovery), which supplies both. We evaluated APOD at three levels. First, we assessed whether the agent recovers a known ground-truth model, using synthetic PK and tumor-dynamics (TD) datasets with realistic sampling and benchmarking it against an ablated version of APOD reasoning only on a scalar metric, and an established automated model-selection tool. Second, we tested it on real experimental data, spanning clinical PK, experimental TD and experimental PK-TD response cohorts. Third, we used it to build a population digital twin for hematotoxicity prediction in radioligand therapy -- an anti-cancer setting for which no myelosuppression model has been established -- and evaluated its forecasts on a held-out cohort.

\section{Results}

\subsection{APOD: Agentic Population ODE Discovery}

We present APOD, a reasoning-guided language-model agent that discovers population mechanistic models (both the shared ODE structural model and the between-subject variability statistical model) from a single prompt (Fig.~1). The agent iteratively proposes candidate models, delegates parameter estimation to a local software specialized in population estimation and refines them by reasoning over goodness-of-fit metrics and aggregated residuals, parametric identifiability and structural and parametric plausibility (see Methods). It returns the selected model together with the reasoning trace behind its choice. No domain-specific knowledge is encoded in the prompts, which favor structural reasoning over retrieval of a published model from training memory. Supplementary Fig.~1 shows a worked example on a PK dataset.

\begin{figure}[bp]
\centering
\includegraphics[width=\textwidth,trim=0 0 0 36bp,clip]{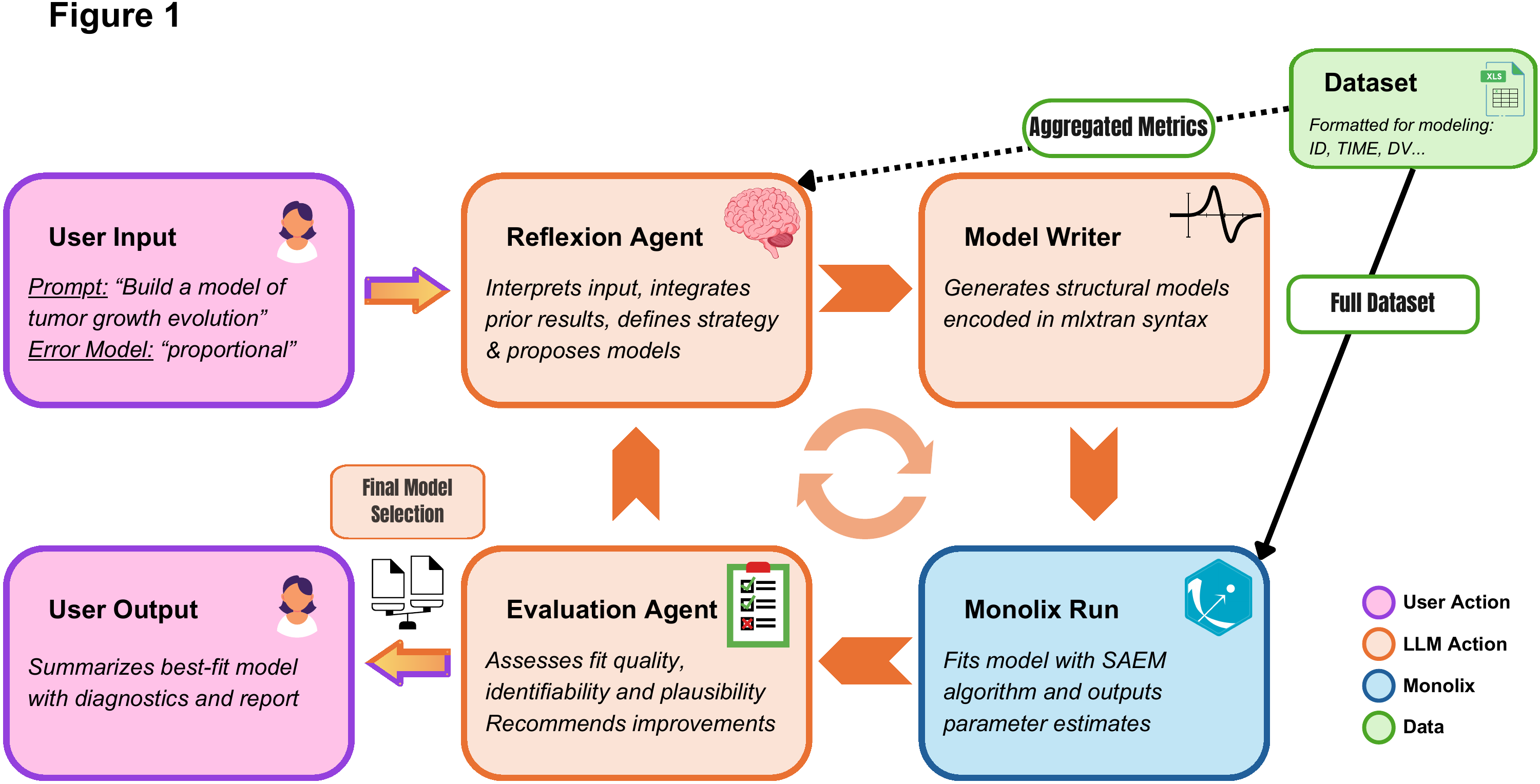}
\caption*{\textbf{Figure 1: The APOD (Agentic Population ODE Discovery) model-discovery workflow.} From a one-sentence task and a residual-error model (here \emph{"Build a model of tumor growth evolution"}, proportional error), the \textbf{reflexion agent} proposes candidate models, each made of a structural ordinary differential equation (ODE) model and a between-subject variability model. The agents see only aggregate metrics of the dataset. The \textbf{model writer} encodes each candidate in Mlxtran, the Monolix modelling language, and \textbf{Monolix} estimates it on the full dataset with the SAEM algorithm. The \textbf{evaluation agent} assesses goodness of fit, parameter identifiability and plausibility, then starts a new iteration (central loop) or stops and returns the selected model with its diagnostics and reasoning trace. Box colours: user actions (purple), language-model actions (orange), Monolix estimation (blue), data (green). LLM: large language model. SAEM: stochastic approximation expectation--maximization. DV: dependent variable.}
\end{figure}

\subsection{APOD recovers ground-truth ODEs by reasoning over fit diagnostics}

We first evaluated the ability of APOD to recover the governing ODE structure in two settings central to pharmacology: pharmacokinetics (PK) and tumor dynamics (TD). We simulated data from ground-truth models under realistic study designs -- sparse sampling, few subjects per arm and realistic measurement noise -- and ran APOD from a minimal prompt (simulation designs and parameters in Supplementary Tables~1-4). Because LLM agents are stochastic, we performed ten independent runs per dataset. We benchmarked APOD against mlxModelFinder, an automated structural-search tool for population PK that searches a predefined library of model structures\textsuperscript{25}.

On the ten synthetic PK datasets, APOD recovered the generating structure in 94\% of runs and mlxModelFinder in 98\%. The set of parameters that vary between subjects was recovered in 66\% and 72\% of runs, respectively. However, APOD's median run was 24 times faster and cost 0.25 US dollars in language-model calls (Table~2).

\begin{sidewaystable}
\landscapepage\centering\captionsetup{font=footnotesize}
\caption*{\textbf{Table 2: APOD recovers ground-truth structures on synthetic pharmacokinetic and tumor-dynamics benchmarks faster than a library-bounded search.} Ten pharmacokinetic (PK) datasets, combining one to three compartments (cmt), intravenous (IV) or oral absorption with or without a lag, and linear or Michaelis--Menten (MM) elimination, and five tumor-dynamics (TD) growth laws (Supplementary Tables 1--4). APOD was run 10 times per dataset and the library search mlxModelFinder (MLX) with 5 estimation seeds. The ablation (5 runs per dataset) selects candidates on the corrected Bayesian information criterion (BICc) alone, without the reasoning layer (Methods). Recovery: percentage of runs recovering the generating structure, then the exact set of random effects as well. Tested models: median number of distinct structures per run or search. Time: median wall-clock time per run. ΔBICc: best APOD run minus best MLX search, a negative value favouring APOD. Cost: median application programming interface cost per run, from the recorded token usage. All datasets: recovery pooled over all runs, other columns as medians. IIV: inter-individual variability. USD: US dollars.}
\includegraphics[width=599bp,height=306bp,trim=0 0 0 17bp,clip]{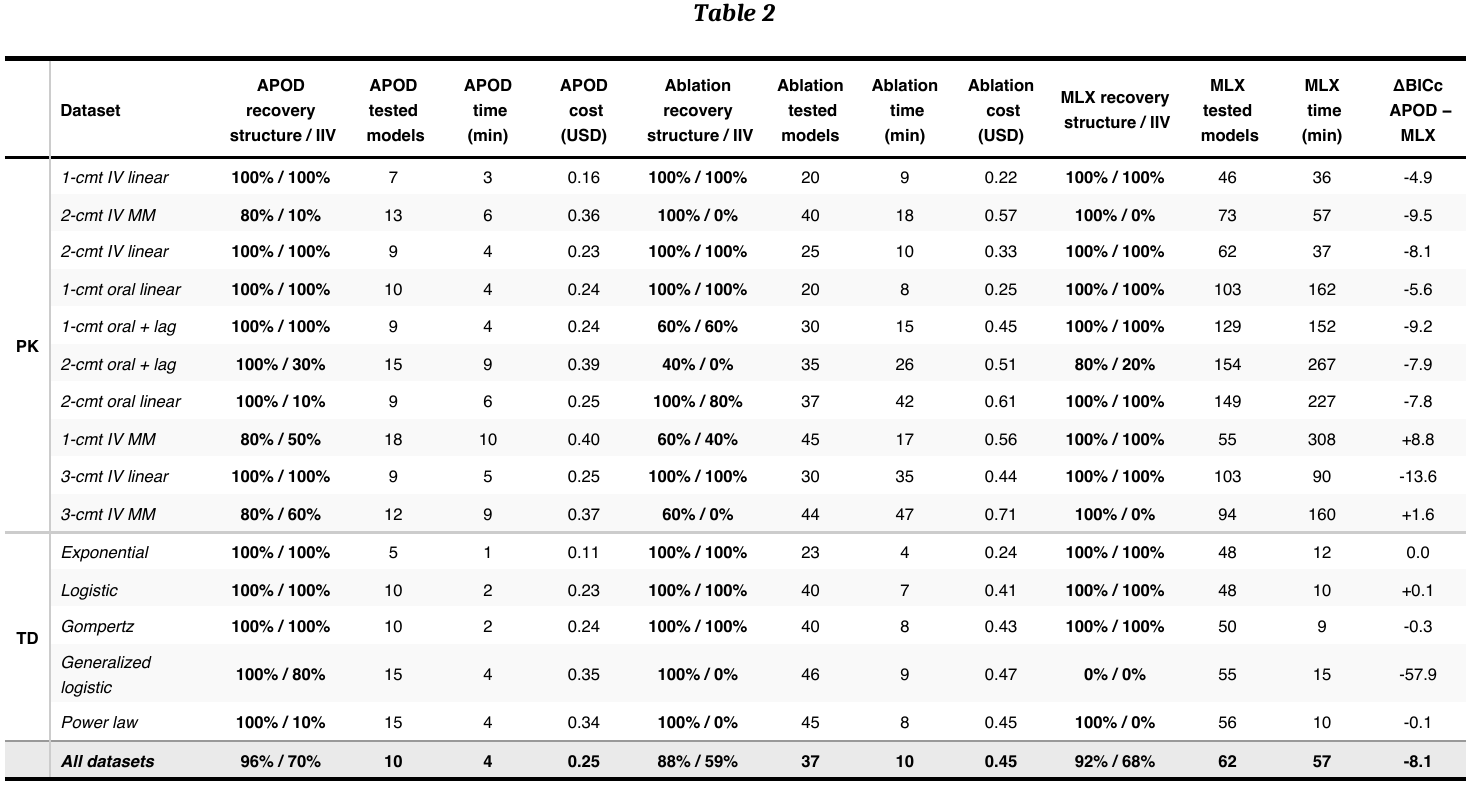}
\end{sidewaystable}

On the five synthetic TD scenarios, APOD recovered the generating structure in 100\% of runs and mlxModelFinder in 80\%, missing the generalized logistic on all five of its seeds. The set of parameters that vary between subjects was recovered in 78\% and 60\% of runs, respectively. APOD's median run was 4.5 times faster and cost 0.24 US dollars in language-model calls (Table~2).

APOD thus recovered the generating structure in 96\% of all runs. Beyond structural recovery, the models selected by APOD provided accurate individual fits across all simulated PK and TD datasets (Supplementary Fig.~2).

APOD's reasoning layer is not specific to a particular model, so we repeated the synthetic benchmark with seven LLMs under identical conditions (Table~3 and Supplementary Table~5). Claude~Opus~4.5, used throughout this work, recovered the generating structure in 96\% of runs, the three other models accessed through an API in 67\% to 89\%, and the three small models run locally in 16\% to 40\% (Table~3).

\begin{sidewaystable}
\landscapepage\centering\captionsetup{font=footnotesize}
\caption*{\textbf{Table 3: Structural recovery across seven language-model backends on the synthetic benchmark.} Each cell gives the percentage of runs recovering the generating structure, then the percentage also recovering the exact set of random effects. A run that did not complete counts as a miss. Each dataset was run 10 times with the reference, Claude Opus 4.5 (bold), and with the three open-weight models run locally through Ollama, and 5 times with the other models, served through an application programming interface (API). All datasets: percentages pooled over all runs. Run time and cost: medians per run, local models incurring no API cost. 20B, 8B, 24B: model size in billions of parameters. Candidates tested, failure rates and run times per dataset are in Supplementary Table 5. PK: pharmacokinetic. TD: tumor dynamics. IIV: inter-individual variability. USD: US dollars.}
\includegraphics[width=472bp,height=326bp,trim=0 0 0 18bp,clip]{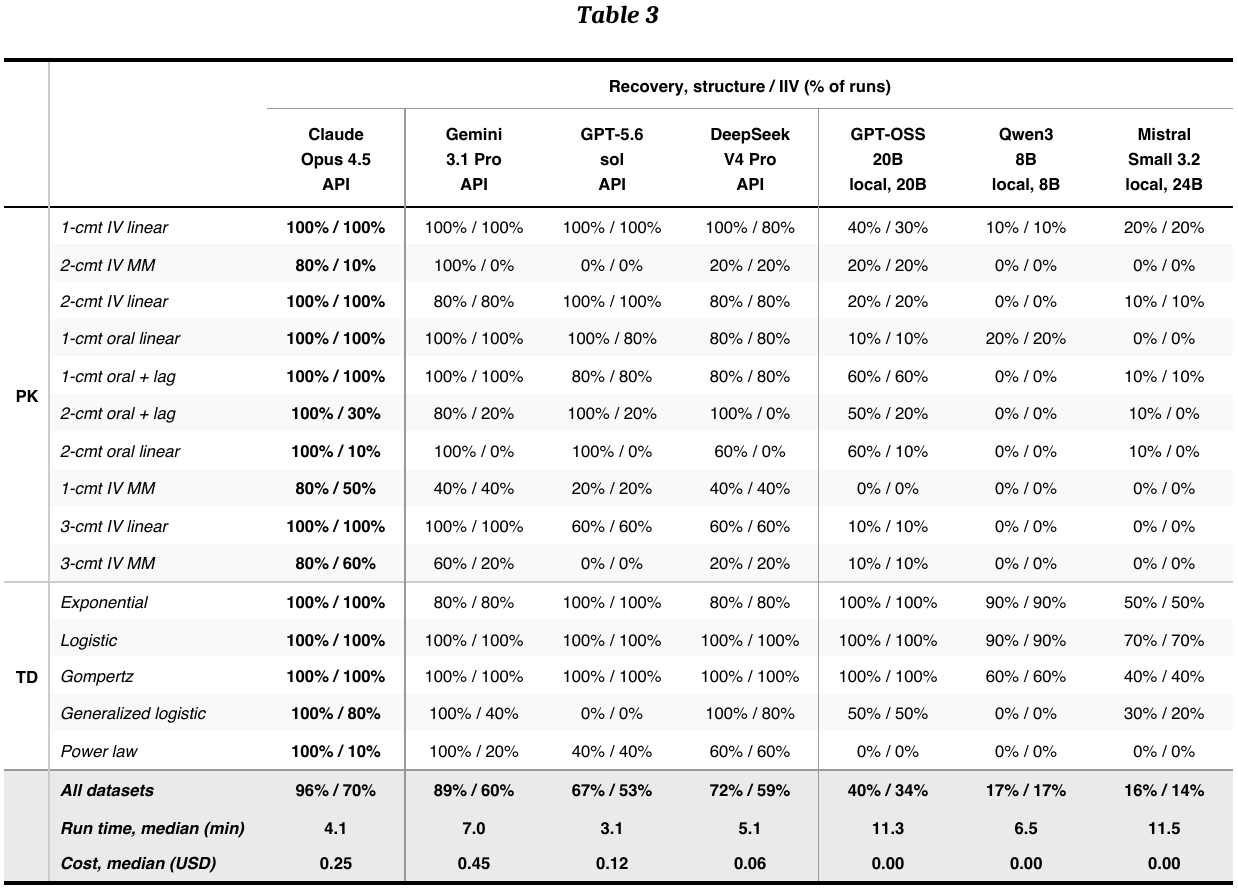}
\end{sidewaystable}

In an ablation experiment without the reasoning layer, the same language model selected candidates on the corrected Bayesian information criterion (BICc) alone (Table~2). Structural recovery fell from 96\% to 88\%, exact random-effect recovery from 70\% to 59\%, and runs took two to three times longer (median 10.8 versus 4.1 min). Within APOD with reasoning, the model retained after the first iteration had the generating structure in 67\% of runs, and the following iterations corrected the first proposal in 44 of the 49 runs where it was wrong.

Together, these results show that APOD recovers correct model structures across PK and TD settings, faster than a library-bounded search.

\subsection{APOD suggests relevant structures on real clinical and experimental cohorts}

\begin{samepage}
On a clinical PK cohort (34 patients, 500 observations), APOD was provided with a drug-agnostic one-sentence prompt -- ``\emph{Build a model for chemotherapeutic drug concentration over time after IV infusion administration}'' (IV, intravenous) -- containing no structural prior (no drug name, compartmental assumption, or elimination mechanism). From this prompt, APOD selected a three-compartment intravenous disposition model consistent with the published analysis\textsuperscript{30}:
\[
\begin{aligned}
\frac{dA_1}{dt} &= -(k_{10}+k_{12}+k_{13}).A_1 + k_{21}.A_2 + k_{31}.A_3\\
\frac{dA_2}{dt} &= k_{12}.A_1 - k_{21}.A_2\\
\frac{dA_3}{dt} &= k_{13}.A_1 - k_{31}.A_3\\
C_c &= \frac{A_1}{V_1},
\end{aligned}
\]
\end{samepage}

\nopagebreak
\noindent with \(C_{c}\ \)the observable of the system (circulating concentration). This structure achieved fit quality comparable to that of the published model\textsuperscript{30}, with identifiable parameters (Fig.~2A, Supplementary Table~6A). Remarkably, APOD was able to suggest a system of three equations to model a single quantity (two latent variables). In a visual predictive check, which compares the observed percentiles over time with those simulated from the model, the observed percentiles fell within their 90\% prediction intervals across the sampling window (Fig.~2B). The same held on the patients held out of discovery (Fig.~2C). Observed and individual-fitted concentrations agreed across the concentration range in those patients (Fig.~2D). It reached similarly plausible, identifiable disposition models on two further real clinical cohorts, oral ciclosporine\textsuperscript{31} and {[}177Lu{]}Lu-DOTATATE renal handling\textsuperscript{32}, each from an equally blinded prompt (Supplementary Figs.~3 and 4).

\begin{figure}[bp]
\centering
\includegraphics[width=\textwidth,trim=0 0 0 0bp,clip]{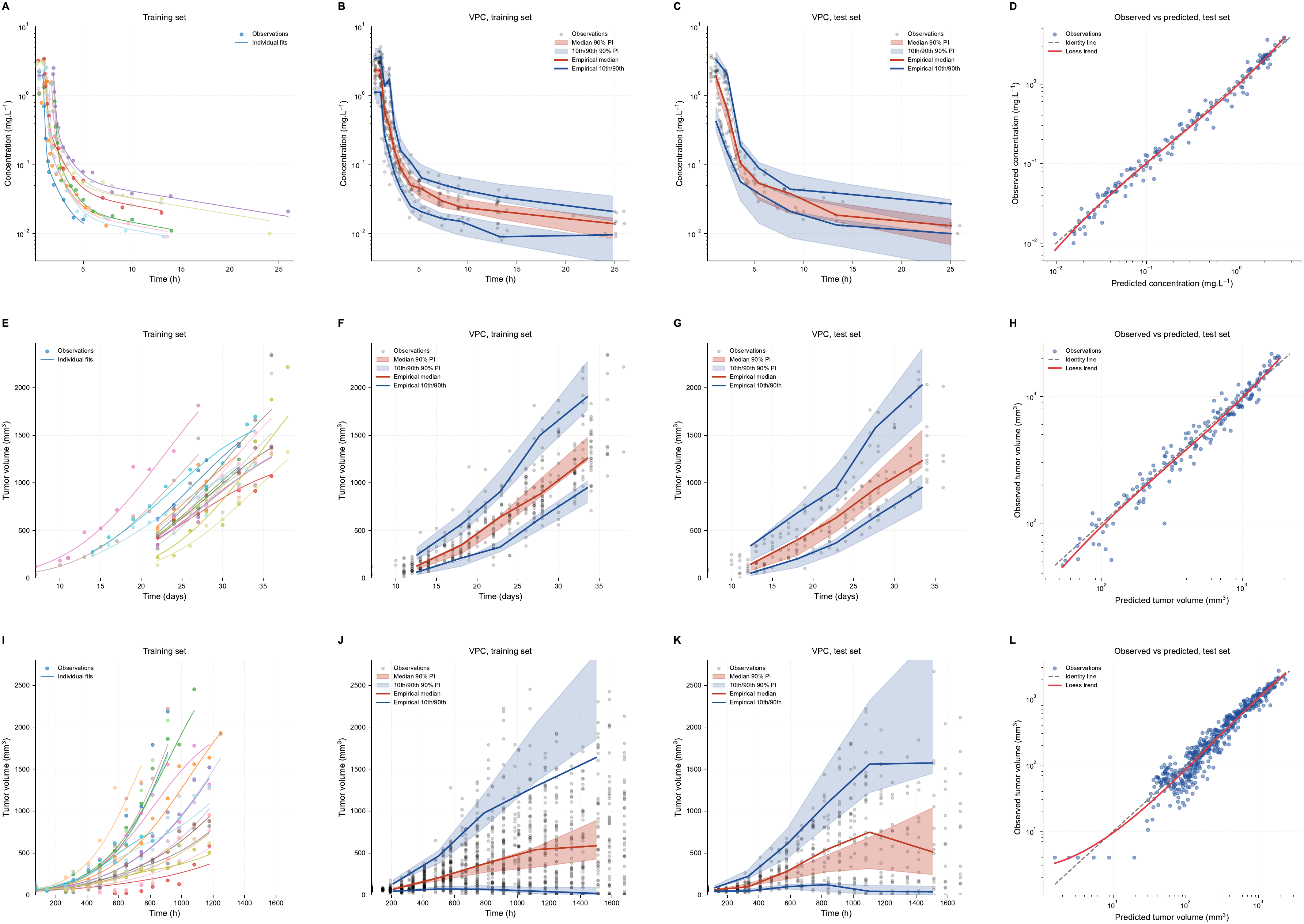}
\caption*{\textbf{Figure 2: APOD models of real clinical and experimental cohorts, and their predictions for held-out subjects.} Each row is one cohort, modelled from a one-sentence prompt naming neither the drug nor a model structure. \textbf{(A--D)} Docetaxel plasma pharmacokinetics\textsuperscript{30}, three-compartment intravenous model, 24 patients for training and 10 held out. \textbf{(E--H)} Untreated tumor growth in breast-cancer xenografts\textsuperscript{33}, Gompertz-type law, 46 mice for training and 20 held out. \textbf{(I--L)} Tumor growth under paclitaxel in MCF-7 breast-cancer xenografts, given as intravenous free drug or as a subcutaneously injectable polymer prodrug\textsuperscript{55}, logistic growth with a threshold drug effect acting through an effect compartment, 74 mice for training and 32 held out. \textbf{(A, E, I)} Individual fits of the first 10 (A) or 20 (E, I) subjects of the training set: observations (points) and individual predictions (lines). \textbf{(B, F, J)} and \textbf{(C, G, K)} Visual predictive checks (VPC) in the training and held-out sets: observed median (red line) and 10th and 90th percentiles (blue lines) against the 90\% prediction intervals (PI) of the same percentiles simulated from the model \textbf{(}bands). \textbf{(D, H, L)} Observed versus individual-predicted values in the held-out set, with the population parameters fixed at their training values, the identity line and a Loess trend. The first three panels of a row share their axes.}
\end{figure}

\begin{samepage}
On the untreated tumor dynamics (TD) cohort\textsuperscript{33} (583 observations, 66 mice), APOD was tasked with identifying a growth model from a minimal description (``\emph{Build a model of tumor growth evolution without treatment effect''}). The agent converged to a hyper-Gompertz growth law with inter-individual variability on two parameters out of four:
\[
\frac{d\mathrm{TS}(t)}{dt} = \lambda.\mathrm{TS}(t).\left[\log\left(\frac{K}{\mathrm{TS}(t)}\right)\right]^{\gamma},
\qquad \mathrm{TS}(t=0) = \mathrm{TS}_0,
\]
\end{samepage}

\nopagebreak
\noindent where \(\mathrm{TS}(t)\) denotes tumor size. The agent generalized the Gompertz model and estimated an exponent well below unity (\(\gamma\) = 0.60, relative standard error, RSE, 1.1\%), a structure outside the usual tumor-growth library. APOD initially placed inter-individual variability (IIV) on all parameters, then removed it on \(\lambda\) and \(\gamma\) after diagnosing high shrinkage, yielding a parsimonious model with all fixed effects at RSE \textless{} 18\% (Supplementary Table~6B). Fits were good at the individual (Fig.~2E) and population (Fig.~2F) levels in the training and test (Fig.~2G-H) sets. Another successful application of APOD to TD data, on a syngeneic lung tumor cohort, is presented in Supplementary Fig.~5.

\begin{samepage}
The next example investigated the ability of APOD to provide a PDT for tumor growth data in mice receiving chemotherapeutic treatment. The data consisted of 11 experimental arms (1780 observations, 106 mice), including intravenous and subcutaneous polymer--prodrug formulations of paclitaxel, across multiple dose levels. The model proposed by APOD combines logistic tumor growth with a drug effect mediated by an effect compartment, the kill term acting only above a concentration threshold:
\[
\begin{aligned}
\frac{d\mathrm{TS}(t)}{dt} = {}& k_g.\mathrm{TS}(t).\left(1-\frac{\mathrm{TS}(t)}{K_{\max}}\right)\\
&- k_d.\max\left(0,\ C_e(t)-C_{\mathrm{th}}\right).\mathrm{TS}(t)
\end{aligned}
\]
\[
\frac{dC_e(t)}{dt} = k_{e0}.\left(C(t)-C_e(t)\right)
\]
\[
\mathrm{TS}(t=0) = \mathrm{TS}_0, \qquad C_e(t=0) = 0,
\]
\end{samepage}

\nopagebreak
\noindent with \(C(t)\) the drug concentration given as an input. Despite the different dose levels and administration schedules, the model accurately predicted the behavior of all mice at both the individual (Fig.~2I) and population (Fig.~2J) levels, and held on the animals excluded from discovery (Fig.~2K-L).

In these real cohorts, APOD built PDTs from structure-agnostic prompts, with identifiable models consistent with those previously published for the corresponding data.

\subsection{APOD provides a PDT that predicts thrombocytopenia under radioligand therapy}

Finally, we turned to a setting with no established model, and tested whether a discovered PDT could be calibrated on new patients from early data and simulated under schedules that were not given.

\begin{samepage}
Radioligand therapy with {[}177Lu{]}Lu-DOTATATE is a standard treatment for metastatic neuroendocrine tumors, in which treatment-induced thrombocytopenia is a frequent toxicity. Individual platelet dynamics have no established mechanistic model\textsuperscript{34}. On the COMPORIV cohort of such patients (Supplementary Table~7)\textsuperscript{35}, from the one-sentence task description alone, APOD converged on the semi-mechanistic myelosuppression Friberg model established for cytotoxic chemotherapy\textsuperscript{36}:
\[
\begin{aligned}
\frac{dA_c}{dt} &= -k_{\mathrm{tr}}.A_c, \qquad E = \mathrm{slope}.A_c\\
\frac{d\,\mathrm{Prol}}{dt} &= k_{\mathrm{tr}}.\mathrm{Prol}.\left(\frac{\mathrm{PLT}_0}{\mathrm{PLT}}\right)^{\gamma}.(1-E) - k_{\mathrm{tr}}.\mathrm{Prol}\\
\frac{d\,\mathrm{Tr}_i}{dt} &= k_{\mathrm{tr}}.(\mathrm{Tr}_{i-1}-\mathrm{Tr}_i), \qquad i=1,\ldots,5\ \ (\mathrm{Tr}_0 \equiv \mathrm{Prol})\\
\frac{d\,\mathrm{PLT}}{dt} &= k_{\mathrm{tr}}.(\mathrm{Tr}_5-\mathrm{PLT}),
\end{aligned}
\]
\end{samepage}

\nopagebreak
\noindent where \(A_{c}\) is a dose-driven drug proxy, \(\mathrm{Prol}\) the proliferating pool, \(\mathrm{Tr}_{1}\)\ldots{}\(\mathrm{Tr}_{5}\) the maturation compartments and \(\mathrm{PLT}\) the circulating platelet count. Although it was given a single observable, the platelet count, APOD introduced a drug-amount proxy \(A_{c}\) driven by the administered dose, an actionable input, and built a system of 7 ODEs with nonlinear feedback for the platelet dynamics. The recovered model described the cohort's platelet-toxicity time course with no sign of misspecification in the individual fits (Supplementary Fig.~6) or the visual predictive checks (Fig.~3A), and with identifiable parameters (Supplementary Table~6D).

\begin{figure}[bp]
\centering
\begin{tikzpicture}\node[inner sep=0,anchor=south west] (img) at (0,0) {\includegraphics[width=\textwidth]{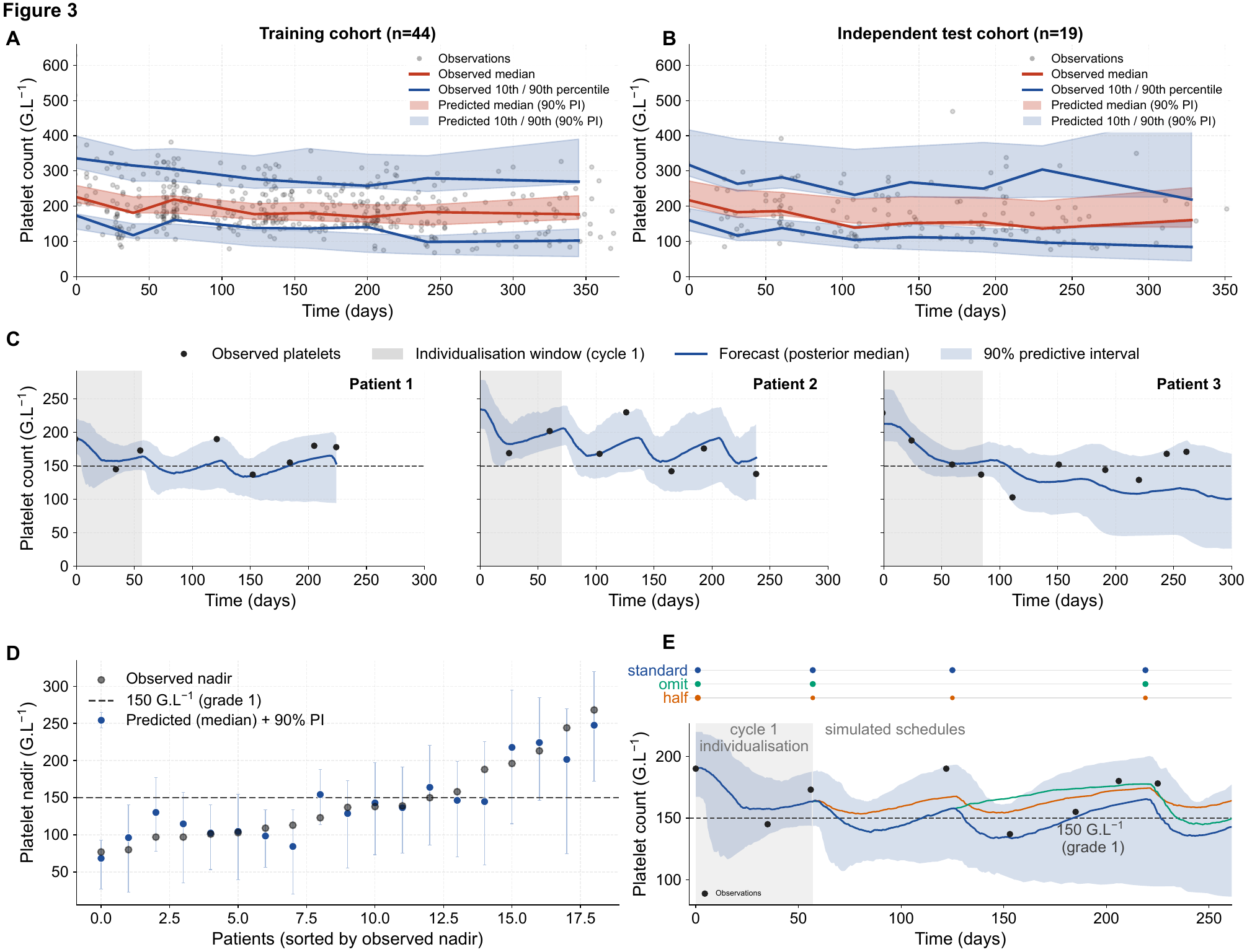}};\begin{scope}[x={(img.south east)},y={(img.north west)}]\fill[white] (-0.0021,0.9741) rectangle (0.0534,1.0048);\fill[white] (0.0497,0.9741) rectangle (0.0664,1.0048);\end{scope}\end{tikzpicture}
\caption*{\textbf{Figure 3: The APOD platelet model as a predictive digital twin of thrombocytopenia under radioligand therapy.} APOD discovered de novo, from a blinded one-sentence prompt, a Friberg-type myelosuppression structure. (A, B) Visual predictive checks (VPC) in the training cohort, on which the model was discovered (n = 44), and in an independent test cohort (n = 19): observed median and 10th and 90th percentiles (lines) against the 90\% prediction intervals (PI) of the same percentiles simulated from the model (bands). (C) First-cycle forecasts for three test patients, with forecast errors of 4\%, 8\% and 22\% (patients 1 to 3, mean absolute relative error): cycle-1 data used for individualisation (grey window), posterior median (line) and 90\% predictive interval (band), later observations (points). (D) Platelet nadir of the 19 test patients, sorted by observed value: observed (grey), median and 90\% PI of the first-cycle forecast (blue), and the 150 G/L threshold of grade-1 thrombocytopenia (dashed). (E) Patient 1, calibrated on cycle 1 and simulated under the schedule given (standard, 7,400 MBq), with one cycle omitted (omit) and with 3,700 MBq from cycle 2 (half). Points: observed counts. Upper rail: dose times, marker size proportional to the administered activity. Band: 90\% PI under the standard schedule.}
\end{figure}

To test whether the discovered structure functions as a predictive model rather than a retrospective fit, we applied it to an independent cohort of 19 patients treated with the same therapy at the same centre over a later period -- a temporal validation\textsuperscript{37,38}. Applied unchanged, the discovered model captured the cohort's platelet distribution at the population level (visual predictive check, Fig.~3B). With the population parameters fixed, each new patient's parameters were estimated by Bayesian individualisation with the population model as prior, from the first treatment cycle alone (a median of three measurements), and we predicted the platelet trajectory over the following cycles and its nadir (lowest count), which governs dose reduction and interruption in this therapy\textsuperscript{34,39}. Fig.~3C shows the first-cycle forecasts for three patients spanning the range of fit quality; all 19 are shown in Supplementary Fig.~7. The 90\% predictive intervals contained the observed nadir in all 19 patients (Fig.~3D). The predicted nadirs agreed with the observed values (concordance correlation 0.91, 95\% confidence interval (CI) 0.79 to 0.96; R² 0.84; Spearman's rank correlation 0.88; root-mean-square error 22 G/L; mean absolute error 17 G/L, 95\% CI 12 to 23). The population model alone, applied to each patient's schedule without individualisation, explained almost none of the between-patient variation in nadir (R² 0.06, root-mean-square error 52 G/L). At the grade-1 thrombocytopenia threshold (150 G/L), this first-cycle forecast flagged 11 of the 12 patients who went on to develop thrombocytopenia and cleared 5 of the 7 who did not.

Because the discovered model is mechanistic with causal dependency on the administered dose, the calibrated twin can be run forward under schedules that were not given. For a patient predicted to cross the grade-1 threshold under the standard schedule, halving the injected activity from the second cycle, the reduction provided for in the label, lowered the predicted probability of a grade-1 nadir from 0.94 to 0.59, whereas omitting one cycle left it at 0.93 (Fig.~3E). Across the 19 patients, the predicted nadir rose by 26 G/L with a halved activity, 7 G/L with an omitted cycle and 2 G/L with a four-week delay of the later cycles.

\section{Discussion}

We have shown that a reasoning-guided language-model agent can discover population digital twins, that is, a shared structural law written as an ODE together with a statistical distribution of its parameters across individuals. Such a model supports adaptive calibration on an individual as measurements arrive and counterfactual simulation of alternative treatments. The novelty is twofold: the population approach for language-model-guided discovery, and a reasoning-guided search that weighs fit diagnostics, parameter identifiability and mechanistic plausibility rather than optimizing a scalar criterion.

Recent work reviewing the field of automated biomedical digital-twin discovery concluded\textsuperscript{12} that no existing methodology simultaneously addresses the eight major biological and methodological challenges posed by real-world biomedical data. APOD addresses six of them simultaneously: (i) sparse and noisy observations, (ii) heterogeneous populations, (iii) incorporation of prior mechanistic knowledge, (iv) candidate-library design, (v) latent variables and (vi) uncertainty quantification, leaving scalability to high-dimensional systems and systematic evaluation under partial observability as important directions for future work. Compared with a library-based automated model-selection tool from the pharmacometrics community\textsuperscript{25}, APOD matched or exceeded structural recovery 2.5- to 40-fold faster, and carries no limitation regarding the application domain. APOD is faster because it fits about six times fewer candidate models than the library search. Importantly, we exemplified this here in two other domains than PK -- the only domain where automated population modeling tools currently exist\textsuperscript{22,24,25} -- namely, TD and pharmacodynamics.

On the synthetic benchmarks the agent navigated a potentially large space of candidate structures and recovered the exact data-generating ODE in most runs, within minutes, under designs reproducing the constraints of real pharmacological data: sparse irregular sampling, few subjects, and unobserved latent compartments. This demonstrates how the integration of prior domain-specific modelling literature can be leveraged by an LLM agent to accelerate model discovery. In addition, the ablation study showed that ranking candidates on BICc alone lowered structural recovery from 96\% to 88\% and made runs two to three times longer, thus showing that reasoning over multiple aspects of a model's adequacy (goodness-of-fit, identifiability and plausibility) outperforms a single-metric approach.

Comparing seven language-model backends, we found that, for the three locally hosted open-weight models, 32\% to 75\% of proposals could not be estimated at all because the generated code was invalid. This may be a limitation for applications where a locally hosted backend is needed but APOD's architecture has been specifically designed to send only aggregated, non-identifying diagnostics to the language model and never individual-level records, thus usable with externally hosted backends.

Structural recovery was consistently better than recovery of the between-subject variability model, most markedly where the simulated variance was small relative to residual error. This is expected, but it is this part of the population model that governs the width of individual predictions, and it is where the method most clearly leaves room for improvement.

On real cohorts no ground truth exists, and APOD was therefore assessed on two criteria: whether the discovered structures are mechanistically interpretable and consistent with published expert analyses, and whether they predict held-out subjects at the population level. Both held across clinical pharmacokinetics and experimental tumour dynamics, from prompts naming neither the drug nor any structural assumption. The radioligand-therapy case tests the more demanding claim. Treatment-induced thrombocytopenia under {[}177Lu{]}Lu-DOTATATE has no established mechanistic model, and from a one-sentence task description the agent converged on a Friberg-type myelosuppression structure\textsuperscript{36}, a class established for cytotoxic chemotherapy and recovered here de novo on a different therapeutic modality. The clinical PK and platelet models contain two and seven never-observed compartments inferred from one measured variable, a step that agents returning an ODE of pre-defined dimension cannot take. Beyond description, the discovered twin supported Bayesian individualisation: from a median of three measurements in the first treatment cycle, its 90\% predictive intervals contained the observed platelet nadir in all 19 patients of a temporally independent cohort. In routine practice, clinicians facing this toxicity delay a cycle, reduce the activity or do both, without evidence to guide the choice\textsuperscript{40}. Because the model is mechanistic, the calibrated twin could then be simulated forward under schedules that were never administered, and the predicted benefit of reducing the activity exceeded that of delaying or omitting a cycle. However, these predictions remain to be properly evaluated in a prospective trial. Together, discovery, individualisation and forward simulation are the sequence a bedside medical DT requires, obtained here from a single cohort where no model existed.

Two limitations bound these results. First, APOD discovers structure and variability but ignores covariates: it explains no part of the between-subject variance by variables available before treatment, such as imaging-derived or genomic features, which a DT needs to predict for a patient with no on-treatment data. Established automatic covariate-building procedures\textsuperscript{41} can be applied sequentially to a discovered model, and the joint discovery of structure and informative variables demonstrated by D3\textsuperscript{17} offers a direct route to bringing that step inside the loop. Second, the clinical validation is temporal rather than multi-centre, observed toxicity was moderate, and prediction of severe dose-limiting events remains to be established on a higher-toxicity cohort.

However, a fundamental question remains. Although the domains examined here (PK, TD, pharmacodynamics) have well-established modeling literature, APOD did not merely retrieve their canonical models: it proposed structures adapted to each dataset, at times converging to functional forms that have been rarely proposed (e.g., a hypergompertz\textsuperscript{42}). What remains untested is discovery in domains that lack any mechanistic description at all. This future avenue will be a decisive test of a reasoning-guided agent: to establish whether APOD can not only recover and adapt known structures, but discover new ones.

\section{Methods}

\subsection{Reasoning-guided discovery agent}

APOD is LLM-agnostic: it can be driven by any cloud-hosted language model through an application programming interface (API) or by a local open-weights model served through Ollama, a tool for running open-weight language models on local hardware (Ollama, version 0.13.4), without changing the discovery loop. For this paper, all experiments were conducted with Claude~Opus~4.5 (OpenRouter model anthropic/claude-opus-4.5). To document this, seven backbones were benchmarked on the synthetic datasets (Table~3; Supplementary Table~5): three proprietary models and one open-weight model accessed through OpenRouter - Claude~Opus~4.5 (anthropic/claude-opus-4.5), GPT-5.6-sol (openai/gpt-5.6-sol), Gemini 3.1 Pro (google/gemini-3.1-pro-preview) and DeepSeek-V4 Pro (deepseek/deepseek-v4-pro) - and three further open-weight models served locally through Ollama : Mistral Small 3.2 (24B, mistral-small3.2:24b), GPT-OSS~20B (gpt-oss:20b) and Qwen3~8B (qwen3:8b). All seven were run through the identical discovery loop, prompts, stopping rules and estimation settings; only the LLM was changed. All other experiments reported in this paper were conducted with Claude~Opus~4.5, which gave the best recovery in this benchmark. As a pinned snapshot, the model carries no persistent memory across runs and performs no test-time or continual learning; every run begins from a fresh context window, so no information leaks between datasets or sessions. The snapshot used throughout corresponds to the Anthropic model identifier claude-opus-4-5-20251101, whose training data extend to August 2025; every Anthropic model identifier denotes a pinned snapshot, so the weights are frozen at that release and identical across all runs reported here.

To preserve data privacy, the language model never received individual-level records: for every dataset it is given only the data structure -- variable names, units, observation types -- together with aggregated initialization metrics computed from the raw data. For the observed variable, these are the cohort baseline (mean of each subject's first measurement) and its mean time, the overall mean across all measurements, and a late-phase value (mean over the 65--85\% time-sorted window) with its mean time. These anchor plausible initial parameter values, and all individual data remain local, used only by the estimation software (Monolix).

For each dataset the agent received a structure-agnostic task description naming the observed quantity but no mechanism or model family. For PK cohorts this was the same single sentence: ``\emph{Build a model for {[}chemotherapeutic{]} drug concentration over time after {[}intravenous infusion / oral{]} administration}''; for tumor dynamics, ``\emph{Build a model of tumor growth evolution}'' (with or without treatment); and for the platelet cohort, \emph{``Build a platelet count model under radioligand therapy''}. The qualifier in brackets was present only for cytotoxic agents and omitted otherwise, and the 177Lu-DOTATATE cohort, whose observation is a renal single-photon emission computed tomography (SPECT) signal rather than a plasma concentration, received a correspondingly adapted sentence.

Supplementary Fig.~1 shows a worked example on a PK dataset. From the user prompt and dataset-level aggregated metrics (Supplementary Fig.~1A), the reflexion agent suggests initial PK models. Given an agnostic template for writing valid Mlxtran code (Supplementary Fig.~1B), the model writer generates valid models, which are fitted locally (Supplementary Fig.~1C) and evaluated for fit, identifiability and plausibility (Supplementary Fig.~1D). This yields a structured judgment that refines the model structure and starts a new loop (Supplementary Fig.~1E). Once enough models have been explored and the LLM is satisfied with its model, the stopping criterion is met and the final model is reported to the user (Supplementary Fig.~1F).

\subsection{Model representation, estimation and iterative refinement}

Each candidate model was expressed as an explicit system of ordinary differential equations in the Mlxtran language. To ensure syntactically valid output, the agent filled a Mlxtran template providing the required block structure, while the equations themselves - and hence the model structure - were left entirely to the agent, so that structures outside any fixed template library could be proposed. The output variable and residual-error model were specified as inputs rather than searched over: for the synthetic benchmarks, the error model used to generate the data; for the real datasets, that of the published reference analysis, or, where none existed, a model specified by the analyst from inspection of the observed data prior to the structural search. Population parameters were estimated by the Stochastic Approximation Expectation--Maximisation (SAEM) algorithm in Monolix, with inter-individual variability (IIV) modelled as log-normal on the structural parameters and estimated with a single Markov chain and automatic stopping disabled. For all runs, the benchmark used 1,000 exploratory and 500 smoothing iterations, with up to four models fitted in parallel. The only exception was the real-world PK-TD (coupled pharmacokinetic and TD) case, which, being more complex, used 1,500 exploratory and 500 smoothing iterations. Standard errors were obtained from the Fisher information matrix estimated by stochastic approximation. The IIV structure was likewise left to the LLM, which could remove variability on individual parameters on the same diagnostic basis. Each agent response was returned as a structured JSON object rather than as free text: the proposal step returns a reasoning field together with a list of candidate models, and the selection step returns the retained model, the reasoning supporting it and a verdict on the residual-error model. These objects are parsed programmatically, so a malformed response is detected rather than silently misread. Regarding software versions, the APOD estimation runs reported here used Monolix Suite 2024R1, whereas the simulation of the synthetic datasets, their identifiability checks and all mlxModelFinder runs used Monolix Suite 2023R1.

After each fit the agent received a compact, non-identifying diagnostic summary comprising: the objective function and information criteria (−2 log-likelihood, Akaike and Bayesian information criteria), per-parameter relative standard errors and η-shrinkage for practical identifiability, individual weighted residuals (IWRES) stratified by time and by variable value, tests of residual normality and bias, and visual-predictive-check scalars (median bias; 90\% predictive interval of the visual predictive check, the percentage of time bins whose observed median lies within the predicted 5th--95th percentile range; and the observed-to-predicted variability ratio, the observed 5th--95th percentile range divided by the predicted one, averaged over bins, which approaches 1 when the model reproduces the observed spread) -- and reasoned over this summary in natural language to decide the next structural move. Each discovery run performed up to ten such structural iterations and could terminate early, the decision being left to the agent's judgement, weighing model quality, convergence and parameter plausibility against the expected gain from further structural exploration. Structures were proposed at a low sampling temperature (0.3) for the initial, literature-anchored model and at a higher temperature (0.7) for subsequent exploratory iterations; malformed proposals were repaired by a small set of syntactic fixes or skipped when invalid.

\subsection{Ablation of the reasoning layer}

To isolate what reasoning over the diagnostics contributes, the discovery loop was re-run with the same model (Claude~Opus~4.5), datasets, prompts for proposing candidates, iteration cap and estimation settings, but with the diagnostic summary reduced to the objective function, the information criteria and the number of parameters, the reasoning and stopping instructions removed, and the model retained at each iteration and at the end taken as the candidate with the lowest BICc. The agent could still end a run by proposing no further candidate. Five runs were performed per dataset and scored as in the benchmark (Table~2).

\subsection{Automated model selection in population modeling with mlxModelFinder}

The library-based autonomous discovery tool mlxModelFinder (Monolix)\textsuperscript{43} was used as a comparator on synthetic datasets. For PK scenarios, the search was performed within the standard PK model library; for TD scenarios, a curated tumor-growth model library was derived from the Monolix library and lightly adapted to retain only structures applicable to tumor growth. Both tools estimated with the same SAEM settings in Monolix (1000 exploratory and 500 smoothing iterations; MonolixSuite 2024R1 for APOD, 2023R1 for mlxModelFinder), with four parallel workers for the PK scenarios and three for tumor dynamics, on the same machine, so that structural search performance and run times are comparable. For the PK scenarios the library was searched in its clearance parameterisation, which can express the independent random effects on clearance and volume of the generative models; the rate parameterisation cannot, and biased selection towards saturable elimination. Each PK search was repeated over five Monolix estimation seeds (the default and 200, 300, 400, 500) under a compute budget of 12 h per search, declared in advance; a search exceeding it counts as a failure, which happened once in 50 searches. For the PK scenarios, mlxModelFinder was run with ant-colony optimisation (population size 20, up to ten iterations), searching the inter-individual-variability structure with the residual error model held fixed, using machine-learning-surrogate scoring (gradient boosting).

\subsection{Synthetic benchmarks}

In the PK context, ten drug-exposure scenarios were constructed to span the main kinetic archetypes encountered in clinical pharmacology, including one-, two- and three-compartment intravenous and oral models with linear or saturable elimination (Supplementary Table~1), parameterized from published analyses of etoposide\textsuperscript{44}, phenytoin\textsuperscript{45}, carboplatin\textsuperscript{46}, theophylline\textsuperscript{47}, imatinib\textsuperscript{48}, tacrolimus\textsuperscript{49}, erlotinib\textsuperscript{50}, cyclophosphamide\textsuperscript{51}, docetaxel\textsuperscript{30,52} and paclitaxel\textsuperscript{53}. Parameter values were taken from the published population analyses of these drugs; the structural families were assigned so that the ten scenarios together span the archetypes of interest, and therefore do not always reproduce the compartmental structure reported in the source analysis. Scenarios were simulated in Simulx (Monolix Suite 2023R1) under a standardized design (per-dataset design in Supplementary Table~2). Each dataset comprised 60 virtual subjects allocated to three dose levels (20 per arm) and sampled at eight to eleven drug-specific nominal times; residual variability used a proportional error model (coefficient 0.15) for every scenario except the three-compartment Michaelis-Menten case (coefficient 0.20) and the three-compartment intravenous linear case, which used a combined error model. Simulations were run with a fixed random seed for reproducibility. Inter-individual variability was applied only to the parameters listed in Supplementary Table~2: the two oral two-compartment scenarios were simulated without variability on Ka (drug\_06 and drug\_07) and on Tlag (drug\_06). These parameters therefore carry no random effect in the generative model; they are not fixed in any estimation sense.

In the TD context, five structural growth models (exponential, gompertz, logistic, generalized logistic, power-law) were used to generate baseline tumor-growth datasets under untreated conditions in a preclinical context (structural models in Supplementary Table~3, simulation design in Supplementary Table~4).

Synthetic datasets were designed to ensure practical parameter identifiability of the data-generating models. For each scenario, we verified that the underlying parameters could be reliably recovered using SAEM when initialized at the ground-truth values. Per-parameter precision was assessed using relative standard error (RSE), and datasets failing this criterion were iteratively revised to ensure recoverability prior to benchmarking.

To assess reproducibility, APOD was run ten times on each dataset.

\subsection{Real-world datasets}

\subsubsection{Real-world model discovery: PK, TD and drug--response coupling}

The first clinical PK cohort for the comparison with classical population modelling was docetaxel phase-I data\textsuperscript{30} (34 patients, 500 observations, a median of 15.5 observations per patient). To probe generalisation across drugs and administration routes, two further real PK cohorts were modelled from blinded, drug-agnostic one-sentence prompts, oral ciclosporine (n~=~14)\textsuperscript{31} and {[}177Lu{]}Lu-DOTATATE renal handling (n~=~10)\textsuperscript{32}; full descriptions and data sources are given in Supplementary Figs.~3 and 4.

The real tumor-growth dataset was taken from a previously published preclinical study of untreated tumor progression in mice\textsuperscript{33,54}. It consists of an orthotopic xenograft model of human triple-negative breast cancer (LM2-4LUC+ cells) with longitudinal caliper-based tumor volume measurements in 66 mice (583 observations). A second tumor-growth dataset, from the same study, was used as an additional real TD application (Supplementary Fig.~5). It consists of a syngeneic subcutaneous Lewis lung carcinoma (LLC) model, with murine LLC cells implanted in 6--8-week-old male C57BL/6 mice, and comprises two independent experiments of 10 animals each (20 mice, 188 observations) followed for approximately 4 to 24 days.

To evaluate the impact of drug exposition on tumor dynamic modelling across diverse dosing regimens, we re-used the polymer--prodrug dataset published in Rodallec et al.\textsuperscript{55}, comprising plasma paclitaxel (Ptx) concentrations and MCF-7 tumor volumes in mice across two administration routes -- intravenous (IV) free Ptx and subcutaneous (SC) Ptx-polyacrylamide (Ptx-PAAm) polymer prodrug. We pooled every arm from the two in vivo campaigns (11 groups, 106 mice, 1780 observations): a model-development campaign (4 groups, 36 mice) with PK sampling at 7 mg/kg eq. Ptx by each route and weekly-dosed efficacy arms (SC PAAm vehicle control at 60 mg/kg eq.; IV Ptx at 15 mg/kg; SC Ptx-PAAm at 15 and 60 mg/kg eq. Ptx), and a schedule-validation campaign (7 groups, 70 mice with SC Ptx-PAAm) spanning total doses of 180, 300, and 540 mg/kg eq. Ptx, from a single bolus and weekly dosing to daily metronomic and loading-dose-plus-maintenance schedules. Rather than re-estimating drug exposure parameters, we reused the population PK model developed and validated in the original study to simulate the individual free Ptx concentration--time profile for every mouse under its own dosing regimen. These simulated exposures, together with the measured tumor volumes across all 106 mice, formed the input to the discovery task: driven by the simulated Ptx exposure, the agent must explain the tumor-volume trajectories alone when discovering the coupled PK-TD model. Spanning routes, doses, and schedules, the dataset thus provides a demanding test of PK-TD discovery under a broad range of dosing regimens.

\subsubsection{Population digital twin construction in clinical radioligand therapy}

We then evaluated APOD using the COMPORIV\textsuperscript{35} cohort from Institut Paoli-Calmettes, consisting of 44 patients with metastatic neuroendocrine tumors treated with {[}177Lu{]}Lu-DOTATATE radioligand therapy, with 519 longitudinal platelet count measurements collected during treatment, a median of 12 per patient. This cohort was used for discovery of radioligand-induced thrombocytopenia dynamics. Patients were included if they had undergone a 68Ga-DOTATOC PET-CT before the first cycle and received at least one cycle of the planned four-cycle course, given eight weeks apart. First cycles were administered between 26 April 2022 and 11 October 2024 in the training cohort, and between 30 October 2024 and 10 November 2025 in the test cohort. Characteristics of both cohorts are given in Supplementary Table~7.

An independent test cohort of 19 patients, treated at the same centre over a later period (temporal validation), was reserved for out-of-sample evaluation. Individual predictions were obtained by Bayesian individualisation from the first treatment cycle with fixed population parameters (see below), enabling patient-specific prediction of platelet trajectories and nadirs without covariates.

\subsection{Evaluation of APOD's discovery and descriptive abilities}

Discovery was assessed differently on synthetic and on real data. On the synthetic benchmarks the data-generating structure was known, so a run was scored as a recovery when the selected model reproduced that structure; the criterion and its per-run reporting are given under Statistics. On the real cohorts no ground truth exists, so the selected models were assessed for descriptive adequacy and for practical identifiability. Descriptive adequacy was judged at the individual level from the observed and individual-predicted profiles and from observed-versus-predicted plots, and at the population level from a visual predictive check, which overlays the observed data on the median and 10th/90th-percentile curves simulated from the fitted population model, each shown with its 90\% prediction interval; a model was considered adequate when the observed percentiles fell within these intervals across the sampling window. Practical identifiability was judged from the relative standard errors of the population parameters and from the eta-shrinkage of the random effects.

\subsection{Held-out evaluation on real datasets}

On the real PK and tumor-dynamics cohorts of at least 20 individuals (patients or mice), discovery was repeated on a training subset and the selected model was evaluated on individuals never seen during discovery. Individuals were assigned at random to a training and a held-out set in a 70/30 ratio with a fixed seed, so that every observation of a given individual falls on one side of the split, and discovery used the same configuration as the full-cohort runs (same residual-error model, same prompt, same selection criteria). Once a model was selected, its population level was frozen entirely (typical values, between-subject variances and residual error) and only the individual random effects of the held-out individuals were estimated. The docetaxel (n~=~34), untreated tumor-growth in murine breast-cancer xenografts (n~=~66 mice), syngeneic lung tumor (n~=~20 mice) and paclitaxel PK-TD (n~=~106 mice) met the threshold, whereas oral ciclosporine (n~=~14) and {[}177Lu{]}Lu-DOTATATE renal handling (n~=~10) did not. For these two cohorts a 30\% held-out set would hold three to four patients, too few for a visual predictive check or a prediction-error estimate to be interpretable, and they are reported on the full cohort.

\subsection{Individual forecasting and predictive validation}

For predictive validation the discovered myelosuppression model and population parameter estimates were fixed to the structure discovered in the COMPORIV cohort. For the test cohort, individual parameters were estimated using only data from the first treatment cycle -- a median of three early platelet measurements per patient -- via maximum a posteriori Bayesian estimation, with the prior distribution derived from the population parameters learned in the training cohort. To derive prediction intervals, samples from the posterior were obtained by a Metropolis--Hastings algorithm using a Gaussian random-walk proposal with step size 0.35 times the population inter-individual standard deviations, initialised at the maximum-a-posteriori estimate, with 300 burn-in iterations and 400 retained draws after thinning by two. Each retained draw was propagated through the model and combined with the residual-error model to give the predictive distribution of the platelet nadir over cycles 2--4, summarised by the median of the simulated nadir draws (rather than the nadir simulated at the posterior mean, which Jensen's inequality would bias for a nonlinear model) and by their 5th--95th percentiles (a 90\% prediction interval). The differential equations were integrated with an adaptive LSODA solver (relative and absolute tolerances 10⁻⁶), and the entire forecasting pipeline was implemented in SciPy independently of Monolix. Counterfactual schedules were simulated from the same posterior draws. For each patient the draws obtained from cycle 1 were propagated a second time after replacing the administered schedule by an alternative one: every cycle after the first delayed by four weeks from the date it was actually given, one cycle omitted, or every cycle from the second halved to 3,700 MBq, the reduction provided for in the label. The structure, the population parameters and the individual draws were left unchanged, so the difference between two arms reflects the dosing schedule alone. These simulations were not compared with observations.

\subsection{Statistics}

Structural recovery on synthetic data is reported as the proportion of ten independent agent runs recovering the data-generating structural family. Recovery of the variability structure counts a run as correct when its selected model carries a random effect on every structural parameter except those simulated without one (Ka and Tlag in the two oral two-compartment scenarios), and is reported over all ten runs, a run that misses the structure counting as a miss. End-to-end run times are medians, over the ten APOD runs and over the five mlxModelFinder searches per PK dataset; a search that exceeded its 12-hour budget counts as a failure in the recovery rates. For the platelet nadir, predictive performance is summarised by the coefficient of determination (R²), Lin's concordance correlation coefficient\textsuperscript{56}, Spearman's rank correlation, and the root-mean-square and mean-absolute errors between predicted and observed nadirs; calibration is the empirical coverage of the 90\% predictive intervals. Applied as a clinical flag, sensitivity and specificity were computed at the grade-1 thrombocytopenia threshold (platelet nadir below 150 G/L). The population-only predictive signal was the R² obtained without individualisation. All statistics were computed in Python (NumPy and SciPy).

\subsection{Computational environment}

All analyses were performed on a single Apple MacBook Pro (M4 Pro system-on-chip, 14 CPU cores, 48 GB unified memory) running macOS 26.5. The discovery agent, diagnostic parsing, Bayesian individualisation and figure generation were implemented in Python 3.10 (NumPy 1.25, SciPy 1.13, pandas 2.2 and Matplotlib 3.9). Population-parameter estimation and data simulation used the Monolix Suite (2024R1 for the APOD estimation runs; 2023R1 for the simulations, the identifiability checks and the mlxModelFinder searches; Monolix for SAEM, Simulx for simulation), driven programmatically through the lixoftConnectors interface under R 4.3.

\subsection{Ethics and data governance}

The COMPORIV training cohort is a sub-cohort of the retrospective, single-centre PIXIRIV study of patients with histologically confirmed neuroendocrine tumors treated with {[}177Lu{]}Lu-DOTATATE at Institut Paoli-Calmettes (Marseille, France), approved by the Institut Paoli-Calmettes institutional ethics committee (approval PIXIRIV-IPC 2023-034) with a waiver of informed consent. The independent test cohort comprises 19 patients treated with the same therapy at the same centre over a subsequent, non-overlapping period, with follow-up extending into 2026. All individual patient data were pseudonymised and processed within a private institutional repository in compliance with the EU General Data Protection Regulation (GDPR), and individual-level records were never exposed to the language model.

\subsection{Use of large language models}

Translation and language editing of the manuscript, and portions of the code development, were assisted by Claude (Anthropic). All outputs were carefully reviewed by the authors.

\subsection{Data availability}

The synthetic benchmark datasets and the discovered Mlxtran model files are available at https://gitlab.inria.fr/compo/papers/APOD\_paper. The preclinical tumor-growth data are available at https://zenodo.org/records/3574531. The clinical cohorts (docetaxel, ciclosporine and the COMPORIV platelet cohorts) contain individual patient data that cannot be shared publicly under the GDPR; they are available from the corresponding author on reasonable request, subject to a data-sharing agreement and institutional approval.

\subsection{Code availability}

The APOD code is available at https://gitlab.inria.fr/compo/apod. The analysis and figure-generation code are available at https://gitlab.inria.fr/compo/papers/APOD\_paper.

\section*{Acknowledgements}\phantomsection\addcontentsline{toc}{section}{Acknowledgements}

This work was supported by a grant from the French government, managed by the National Research Agency (ANR), under the France 2030 program within the DIGPHAT project, reference ANR-22-PESN-0017.

\section*{Author contributions}\phantomsection\addcontentsline{toc}{section}{Author contributions}

R.F. and M.S. contributed equally. S.B. conceived and supervised the study and contributed to writing the manuscript. R.F. and M.S. developed APOD, designed and ran the benchmarks and analyses, and wrote the manuscript, R.F. for the PK and platelet parts and M.S. for the tumor-dynamics parts. V.G. contributed to the early conception of the project. A.M., J.C. and S.O.-T. provided the radioligand-therapy clinical data and clinical expertise. J.N. provided the paclitaxel polymer-prodrug data. F.G. contributed pharmacometric expertise. M.v.d.S. contributed to the methodological design and interpretation. All authors reviewed and approved the manuscript.

\section*{Competing interests}\phantomsection\addcontentsline{toc}{section}{Competing interests}

The authors declare no competing interests.

\section*{References}
\begingroup\referencestyle\phantomsection\addcontentsline{toc}{section}{References}

1. National Academy of Engineering \& National Academies of Sciences, E., and Medicine. \emph{Foundational Research Gaps and Future Directions for Digital Twins}. (The National Academies Press, Washington, DC, 2024). doi:10.17226/26894.

2. Laubenbacher, R., Mehrad, B., Shmulevich, I. \& Trayanova, N. Digital twins in medicine. \emph{Nat. Comput. Sci.} \textbf{4}, 184--191 (2024).

3. Sadée, C. \emph{et al.} Medical digital twins: enabling precision medicine and medical artificial intelligence. \emph{Lancet Digit. Health} \textbf{7}, 100864 (2025).

4. Olawade, D. B. \emph{et al.} Digital twins in oncology: From predictive modelling to personalised treatment strategies. \emph{Crit. Rev. Oncol. Hematol.} \textbf{220}, 105171 (2026).

5. Hernandez-Boussard, T. \emph{et al.} Digital twins for predictive oncology will be a paradigm shift for precision cancer care. \emph{Nat. Med.} \textbf{27}, 2065--2066 (2021).

6. Willcox, K. \& Segundo, B. The role of computational science in digital twins. \emph{Nat. Comput. Sci.} \textbf{4}, 147--149 (2024).

7. Darwich, A. S. \emph{et al.} Model-Informed Precision Dosing: Background, Requirements, Validation, Implementation, and Forward Trajectory of Individualizing Drug Therapy. \emph{Annu. Rev. Pharmacol. Toxicol.} \textbf{61}, 225--245 (2021).

8. Marshall, S. F. \emph{et al.} Good Practices in Model-Informed Drug Discovery and Development: Practice, Application, and Documentation. \emph{CPT Pharmacomet. Syst. Pharmacol.} \textbf{5}, 93--122 (2016).

9. Schmidt, M. \& Lipson, H. Distilling Free-Form Natural Laws from Experimental Data. \emph{Science} \textbf{324}, 81--85 (2009).

10. Cranmer, M. Interpretable Machine Learning for Science with PySR and SymbolicRegression.jl. Preprint at https://doi.org/10.48550/arXiv.2305.01582 (2023).

11. Brunton, S. L., Proctor, J. L. \& Kutz, J. N. Discovering governing equations from data by sparse identification of nonlinear dynamical systems. \emph{Proc. Natl. Acad. Sci.} \textbf{113}, 3932--3937 (2016).

12. Métayer, C., Ballesta, A. \& Martinelli, J. Data-driven discovery of digital twins in biomedical research. \emph{Brief. Bioinform.} \textbf{27}, bbaf722 (2026).

13. Wang, H. \emph{et al.} Scientific discovery in the age of artificial intelligence. \emph{Nature} \textbf{620}, 47--60 (2023).

14. Romera-Paredes, B. \emph{et al.} Mathematical discoveries from program search with large language models. \emph{Nature} \textbf{625}, 468--475 (2024).

15. Holt, S., Liu, T. \& Schaar, M. V. D. Automatically Learning Hybrid Digital Twins of Dynamical Systems. in \emph{Advances in Neural Information Processing Systems 37} 72170--72218 (Neural Information Processing Systems Foundation, Inc. (NeurIPS), Vancouver, BC, Canada, 2024). doi:10.52202/079017-2304.

16. Shojaee, P. \emph{et al.} LLM-SRBench: A New Benchmark for Scientific Equation Discovery with Large Language Models. Preprint at https://doi.org/10.48550/arXiv.2504.10415 (2025).

17. Holt, S., Qian, Z., Liu, T., Weatherall, J. \& Schaar, M. V. D. Data-Driven Discovery of Dynamical Systems in Pharmacology using Large Language Models. in \emph{Advances in Neural Information Processing Systems 37} 96325--96366 (Neural Information Processing Systems Foundation, Inc. (NeurIPS), Vancouver, BC, Canada, 2024). doi:10.52202/079017-3053.

18. Sheiner, L. B., Rosenberg, B. \& Melmon, K. L. Modelling of individual pharmacokinetics for computer-aided drug dosage. \emph{Comput. Biomed. Res.} \textbf{5}, 441--459 (1972).

19. Lavielle, M. \emph{Mixed Effects Models for the Population Approach}. (Chapman and Hall/CRC, 2014). doi:10.1201/b17203.

20. Virgolin, M. \& Pissis, S. Symbolic regression is NP-hard. \emph{Trans. Mach. Learn. Res.} \textbf{10}, 1--11 (2022).

21. Bies, R. R. \emph{et al.} A Genetic Algorithm-Based, Hybrid Machine Learning Approach to Model Selection. \emph{J. Pharmacokinet. Pharmacodyn.} \textbf{33}, 195--221 (2006).

22. Li, X. \emph{et al.} pyDarwin machine learning algorithms application and comparison in nonlinear mixed-effect model selection and optimization. \emph{J. Pharmacokinet. Pharmacodyn.} \textbf{51}, 785--796 (2024).

23. Richardson, S. \emph{et al.} A machine learning approach to population pharmacokinetic modelling automation. \emph{Commun. Med.} \textbf{5}, 327 (2025).

24. Chen, X. \emph{et al.} A fully automatic tool for development of population pharmacokinetic models. \emph{CPT Pharmacomet. Syst. Pharmacol.} \textbf{13}, 1784--1797 (2024).

25. Pierre, M., Mihaljevic, F., Celliere, G. \& Bertrand, J. Implementation and comparison of six algorithms for automated model selection: two simulation studies and eleven real case studies. SSRN Scholarly Paper at https://doi.org/10.2139/ssrn.5956863 (2025).

26. Hornik, K., Stinchcombe, M. \& White, H. Multilayer feedforward networks are universal approximators. \emph{Neural Netw.} \textbf{2}, 359--366 (1989).

27. Chen, R. T. Q., Rubanova, Y., Bettencourt, J. \& Duvenaud, D. K. Neural ordinary differential equations. in \emph{Advances in neural information processing systems} (eds Bengio, S. et al.) vol. 31 (Curran Associates, Inc., 2018).

28. Lu, J., Bender, B., Jin, J. Y. \& Guan, Y. Deep learning prediction of patient response time course from early data via neural-pharmacokinetic/pharmacodynamic modelling. \emph{Nat. Mach. Intell.} \textbf{3}, 696--704 (2021).

29. Laurie, M. \& Lu, J. Explainable deep learning for tumor dynamic modeling and overall survival prediction using Neural-ODE. \emph{Npj Syst. Biol. Appl.} \textbf{9}, 58 (2023).

30. Launay-Iliadis, M. C. \emph{et al.} Population pharmacokinetics of docetaxel during phase I studies using nonlinear mixed-effect modeling and nonparametric maximum-likelihood estimation. \emph{Cancer Chemother. Pharmacol.} \textbf{37}, 47--54 (1995).

31. Caroli‐Bosc, F. \emph{et al.} Ursodeoxycholic acid modulates cyclosporin A oral absorption in liver transplant recipients*. \emph{Fundam. Clin. Pharmacol.} \textbf{14}, 601--609 (2000).

32. Devasia, T. P., Dewaraja, Y. K., Frey, K. A., Wong, K. K. \& Schipper, M. J. A Novel Time--Activity Information-Sharing Approach Using Nonlinear Mixed Models for Patient-Specific Dosimetry with Reduced Imaging Time Points: Application in SPECT/CT After 177Lu-DOTATATE. \emph{J. Nucl. Med.} \textbf{62}, 1118--1125 (2021).

33. Vaghi, C. \emph{et al.} Population modeling of tumor growth curves and the reduced Gompertz model improve prediction of the age of experimental tumors. \emph{PLOS Comput. Biol.} \textbf{16}, e1007178 (2020).

34. Ferrara, R. \emph{et al.} Predictive factors for persistent thrombocytopenia after peptide receptor radioligand therapy in enteropancreatic neuroendocrine tumors. \emph{Front. Endocrinol.} \textbf{16}, (2025).

35. Ferrara, R. \emph{et al.} Deep Learning-Based Whole-Body Lesion Segmentation and Automated OMIS Computation on {[}68Ga{]}Ga-DOTA-TOC PET/CT: Technical Feasibility of a SwinUNETR Pipeline for Pre-PRRT Bone Marrow Involvement Scoring. Preprint at https://doi.org/10.21203/rs.3.rs-10101940/v1 (2026).

36. Friberg, L. E., Henningsson, A., Maas, H., Nguyen, L. \& Karlsson, M. O. Model of Chemotherapy-Induced Myelosuppression With Parameter Consistency Across Drugs. \emph{J. Clin. Oncol.} \textbf{20}, 4713--4721 (2002).

37. Collins, G. S., Reitsma, J. B., Altman, D. G. \& Moons, K. G. M. Transparent reporting of a multivariable prediction model for individual prognosis or diagnosis (TRIPOD): the TRIPOD statement. \emph{BMJ} \textbf{350}, g7594 (2015).

38. Altman, D. G. \& Royston, P. What do we mean by validating a prognostic model? \emph{Stat. Med.} \textbf{19}, 453--473 (2000).

39. Lutathera \textbar{} European Medicines Agency (EMA). https://www.ema.europa.eu/en/medicines/human/EPAR/lutathera (2018).

40. Oziel-Taieb, S., Ansquer, C., Tlili, G., Deshayes, E. \& Haissaguerre, M. Managing hematological toxicity in {[}\textsuperscript{177}Lu{]}Lu-DOTATATE for NET patients: a national survey from the French group of endocrine tumors (GTE) on the clinical need for harmonized monitoring strategies. \emph{Eur. J. Nucl. Med. Mol. Imaging} https://doi.org/10.1007/s00259-026-08118-4 (2026) doi:10.1007/s00259-026-08118-4.

41. Karlsen, M. \emph{et al.} Covariate Model Selection Approaches for Population Pharmacokinetics: A Systematic Review of Existing Methods, From SCM to AI. \emph{CPT Pharmacomet. Syst. Pharmacol.} \textbf{14}, 621--639 (2025).

42. Turner, M. E., Bradley, E. L., Kirk, K. A. \& Pruitt, K. M. A theory of growth. \emph{Math. Biosci.} \textbf{29}, 367--373 (1976).

43. Package `mlxModelFinder' \textbar{} MonolixSuite Documentation. https://monolixsuite.slp-software.com/r-functions/2024R1/package-mlxmodelfinder.

44. Würthwein, G. \emph{et al.} Population pharmacokinetics of high-dose etoposide in children receiving different conditioning regimens. \emph{Anticancer. Drugs} \textbf{13}, 101--110 (2002).

45. Sheiner, L. B. \& Beal, S. L. Evaluation of methods for estimating population pharmacokinetic parameters. I. Michaelis-menten model: Routine clinical pharmacokinetic data. \emph{J. Pharmacokinet. Biopharm.} \textbf{8}, 553--571 (1980).

46. Chatelut, E. \emph{et al.} Prediction of Carboplatin Clearance From Standard Morphological and Biological Patient Characteristics. \emph{JNCI J. Natl. Cancer Inst.} \textbf{87}, 573--580 (1995).

47. Boeckmann, A. J., Sheiner, L. B., \& Beal, S. L. \emph{NONMEM Users Guide: Part V, Introductory Guide}. (1994).

48. Schmidli, H. \emph{et al.} Population pharmacokinetics of imatinib mesylate in patients with chronic-phase chronic myeloid leukaemia: results of a phase III study. \emph{Br. J. Clin. Pharmacol.} \textbf{60}, 35--44 (2005).

49. Staatz, C. E. \& Tett, S. E. Clinical pharmacokinetics and pharmacodynamics of tacrolimus in solid organ transplantation. \emph{Clin. Pharmacokinet.} \textbf{43}, 623--653 (2004).

50. Lu, J.-F. \emph{et al.} Clinical pharmacokinetics of erlotinib in patients with solid tumors and exposure-safety relationship in patients with non-small cell lung cancer. \emph{Clin. Pharmacol. Ther.} \textbf{80}, 136--145 (2006).

51. McCune, J. S. \emph{et al.} Population Pharmacokinetics of Cyclophosphamide and Metabolites in Children With Neuroblastoma: A Report From the Children's Oncology Group. \emph{J. Clin. Pharmacol.} \textbf{49}, 88--102 (2009).

52. Bruno, R. \emph{et al.} A population pharmacokinetic model for docetaxel (Taxotere®): Model building and validation. \emph{J. Pharmacokinet. Biopharm.} \textbf{24}, 153--172 (1996).

53. Henningsson, A. \emph{et al.} Mechanism-Based Pharmacokinetic Model for Paclitaxel. \emph{J. Clin. Oncol.} \textbf{19}, 4065--4073 (2001).

54. Benzekry, S. \emph{et al.} Classical Mathematical Models for Description and Prediction of Experimental Tumor Growth. \emph{PLOS Comput. Biol.} \textbf{10}, e1003800 (2014).

55. Rodallec, A. \emph{et al.} Model-Driven Optimization of Subcutaneous Polymer Prodrugs Achieves Cancer Remission in Mice. \emph{Adv. Healthc. Mater.} \textbf{15}, e05872 (2026).

56. Lin, L. I.-K. A Concordance Correlation Coefficient to Evaluate Reproducibility. \emph{Biometrics} \textbf{45}, 255--268 (1989).

\endgroup

\clearpage


\section*{Supplementary material}

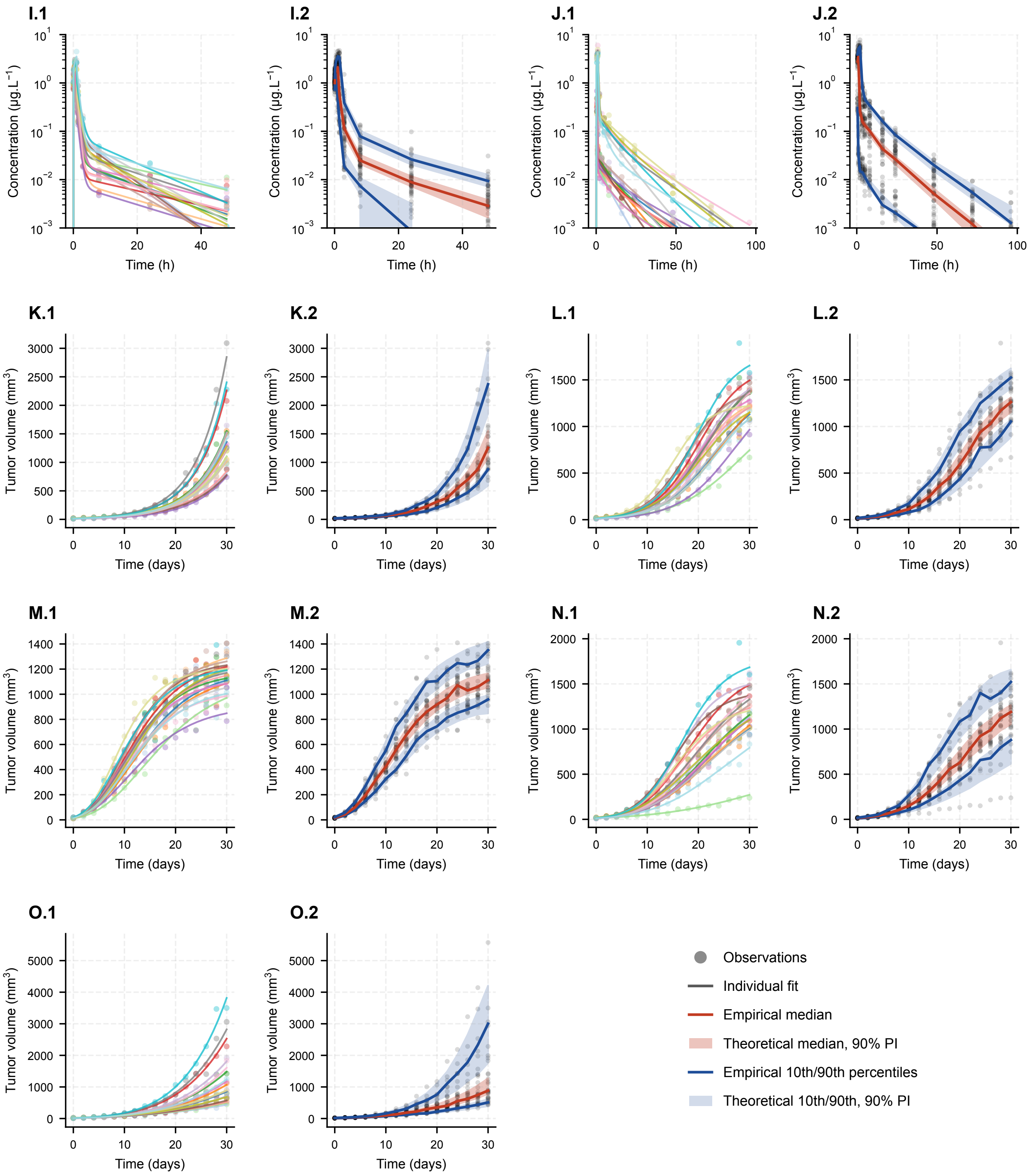


**Legend.** Individual fits and visual predictive checks (VPC) for every recovered synthetic pharmacokinetic (PK) and tumor-dynamics (TD) benchmark archetype (Supplementary Tables 1–4); each letter is one archetype, its `.1` panel showing individual fits (per-subject colors are arbitrary) and its `.2` panel the corresponding VPC – color key above. PI: prediction interval; cmt: compartment. A: 1-cmt IV (intravenous), linear; B: 2-cmt IV, Michaelis-Menten; C: 2-cmt IV, linear; D: 1-cmt oral, linear; E: 1-cmt oral, lag; F: 2-cmt oral, lag; G: 2-cmt oral, linear; H: 1-cmt IV, Michaelis-Menten; I: 3-cmt IV, linear; J: 3-cmt IV, Michaelis-Menten; K: Exponential; L: Logistic; M: Gompertz; N: Generalized logistic; O: Power law.

# Supplementary Figure 3: Real pharmacokinetic cohort — ciclosporine (oral)

**A**

```
[LONGITUDINAL]
input = {ka, gamma, V1, V2, Q, ke}

PK:
depot(target=Ad)

EQUATION:
t_0 = 0
Ad_0 = 0
Ac_0 = 0
Ap_0 = 0

; Weibull absorption: rate increases with time when gamma > 1
if t > 0
    weibull_rate = ka * gamma * (ka*t)^(gamma-1)
else
    weibull_rate = ka
end

ddt_Ad = -weibull_rate * Ad
ddt_Ac = weibull_rate * Ad - ke * Ac - (Q/V1) * Ac + (Q/V2) * Ap
ddt_Ap = (Q/V1) * Ac - (Q/V2) * Ap

Cc = Ac / V1

OUTPUT:
output = Cc
```

**B**

| Parameter | Unit | Value | RSE (%) | 95% CI |
|---|---|---|---|---|
| *ka_pop* | $h^{-1}$ | 0.663 | 15.9 | [0.488, 0.9] |
| *gamma_pop* | — | 1.48 | 7.52 | [1.28, 1.71] |
| *V1_pop* | L | 65.8 | 14.9 | [49.4, 87.7] |
| *V2_pop* | L | 116 | 21 | [77.7, 172] |
| *Q_pop* | $L.h^{-1}$ | 33.2 | 12.8 | [25.9, 42.5] |
| *ke_pop* | $h^{-1}$ | 0.391 | 16.3 | [0.286, 0.535] |
| *omega_ka* | — | 0.506 | 21.4 | [0.337, 0.76] |
| *omega_gamma* | — | 0.246 | 25.8 | [0.152, 0.399] |
| *omega_V1* | — | 0.24 | 21.9 | [0.159, 0.363] |
| *omega_V2* | — | 0.0808 | 244 | [0.0104, 0.629] |
| *b* | — | 0.213 | 7.09 | [0.186, 0.245] |

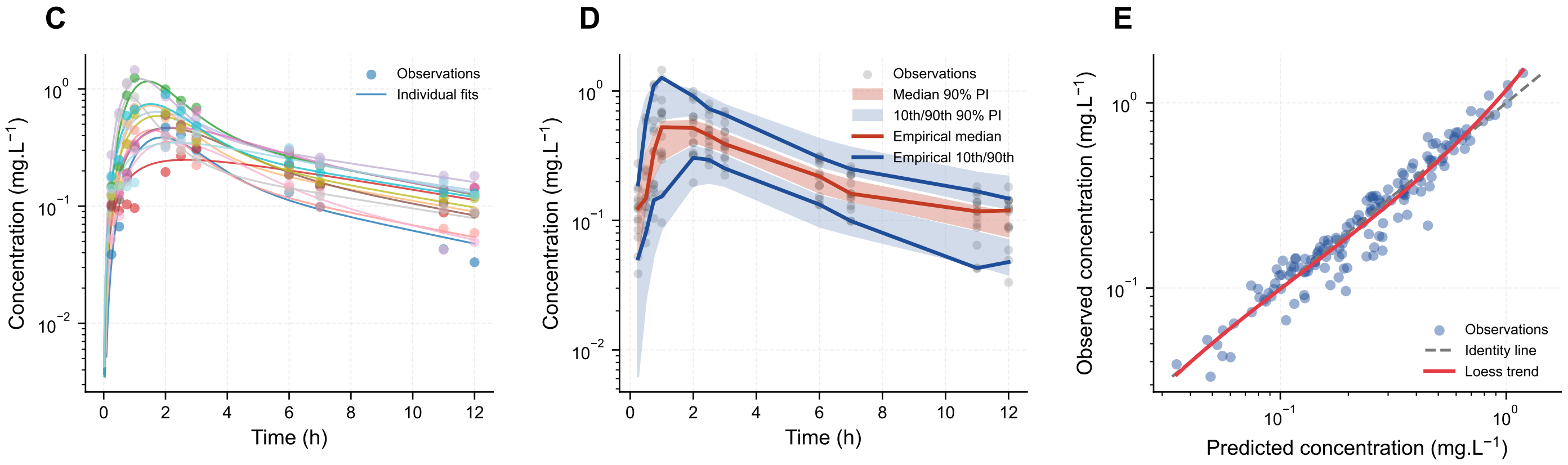


**Legend.** APOD-discovered model for the real ciclosporine (oral, Neoral) pharmacokinetic cohort ($n = 14$, 154 observations), obtained from a blinded, drug-agnostic one-sentence prompt. (A) Selected Mlxtran model code; $k_a$, absorption rate constant; $\gamma$, Weibull shape parameter; $V_1$, $V_2$, central and peripheral volumes; $Q$, inter-compartmental clearance; $k_e$, elimination rate constant; $\omega$, standard deviation of the log-normal random effect; $b$, proportional residual error. (B) Population parameter estimates with relative standard errors (RSE) and 95% confidence intervals (CI) from SAEM (stochastic approximation expectation–maximization) estimation. (C) Individual fits: observed (points) and individual-predicted (lines) concentrations. (D) Visual predictive check: observed data against the model-predicted median and 10th/90th percentile intervals (PI, prediction interval). (E) Observed versus individual-predicted concentrations, with the identity line and a Loess trend; Supplementary Figure 4 follows the same layout. With fewer than 30 patients this cohort was not split into training and held-out sets: a 30% held-out set would hold 4 patients, too few for a visual predictive check or a prediction-error estimate to be interpretable, so the model is reported on the full cohort (Methods). Data source: Caroli-Bosc et al., Fundam. Clin. Pharmacol. 14:601 (2000).

# Supplementary Figure 4: Real pharmacokinetic cohort — $^{177}$Lu-DOTATATE (Lutathera, renal handling)

**A**

```
[LONGITUDINAL]
input = {V, k}

PK:
depot(target=A)

EQUATION:
t_0 = 0
A_0 = 0
ddt_A = -k*A
C = A/V

OUTPUT:
output = {C}
```

**B**

| Parameter | Unit | Value | RSE (%) | 95% CI |
|---|---|---|---|---|
| *V_pop* | — | 91.2 | 8.33 | [77.5, 107] |
| *k_pop* | $h^{-1}$ | 0.0112 | 8.19 | [0.00956, 0.0132] |
| *omega_V* | — | 0.214 | 31.9 | [0.119, 0.384] |
| *omega_k* | — | 0.204 | 35.1 | [0.108, 0.386] |
| *b* | — | 0.18 | 17.5 | [0.128, 0.252] |

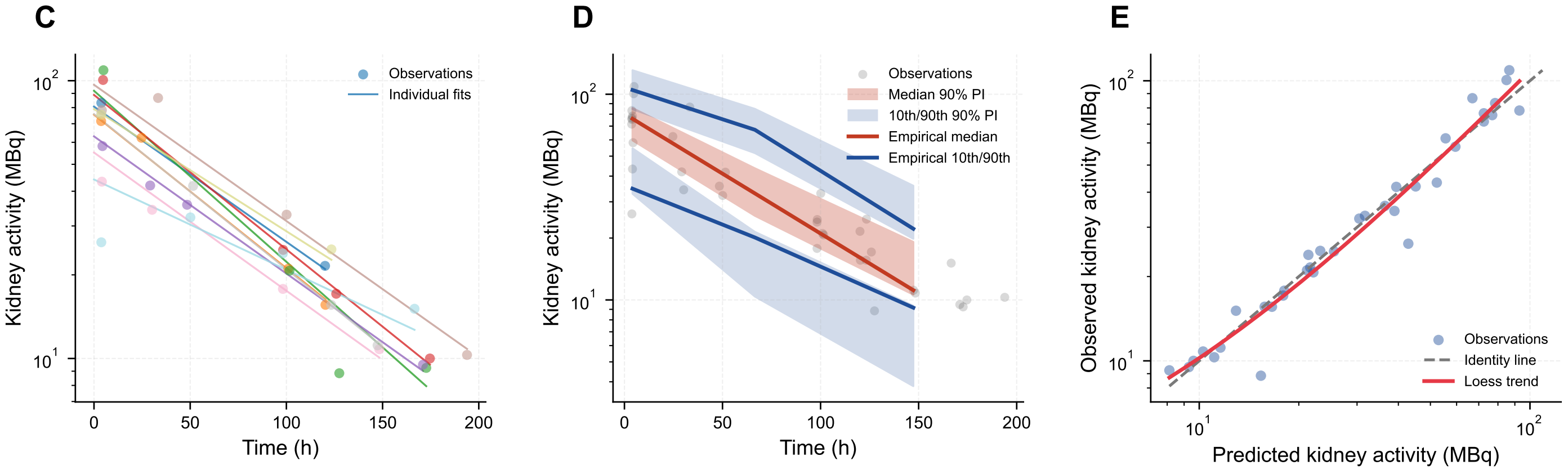


**Legend.** APOD-discovered model for the real $^{177}$Lu-DOTATATE (Lutathera) renal-handling cohort ($n = 10$ patients), from a blinded prompt; panels A–E as in Supplementary Figure 3 (RSE, relative standard error; CI, confidence interval; PI, prediction interval; $k$, elimination rate constant; $\omega$, standard deviation of the log-normal random effect; $b$, proportional residual error). With fewer than 30 patients this cohort was not split into training and held-out sets: a 30% held-out set would hold 3 patients, too few for a visual predictive check or a prediction-error estimate to be interpretable, so the model is reported on the full cohort (Methods). Kidney activity is in MBq and time in hours; the injected activity is also in MBq, so the apparent volume $V$ is a dimensionless dilution factor – the fitted value of 91 corresponds to 1.1% of the injected activity reaching the kidney, consistent with the reported renal uptake for this therapy. Data source: Devasia et al., J. Nucl. Med. 62:1118 (2021).

# Supplementary Figure 5: Real TD lung cohort

**A**

```
[LONGITUDINAL]
input = {k, V0}

EQUATION:
t_0 = 0
K_fixed = 1800
V_0 = V0
ddt_V = k*V*(1 - V/K_fixed)

OUTPUT:
output = {V}
```

**B**

| Parameter | Unit | Value | RSE (%) | 95% CI |
|---|---|---|---|---|
| *k_pop* | $day^{-1}$ | 0.374 | 4.9 | [0.34, 0.411] |
| *V0_pop* | $mm^3$ | 3.43 | 25.8 | [2.11, 5.56] |
| *omega_k* | — | 0.166 | 23.9 | [0.106, 0.261] |
| *omega_V0* | — | 0.841 | 30.7 | [0.477, 1.48] |
| *b* | — | 0.191 | 7.54 | [0.165, 0.221] |

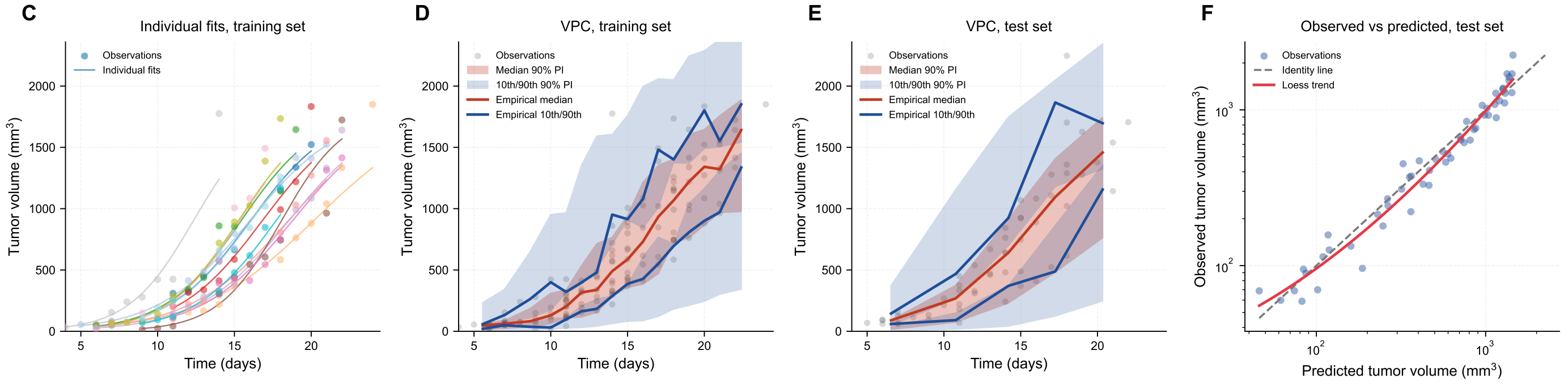


**Legend.** APOD-discovered model for a real lung tumor-volume cohort (tumor dynamics, TD), fit on a train/test split. Data are from a syngeneic subcutaneous Lewis lung carcinoma (LLC) model (murine LLC cells implanted in 6–8 week-old male C57BL/6 mice), comprising two independent experiments of 10 animals each: 14 mice (133 observations) for estimation, 6 held-out mice (55 observations) for evaluation, over follow-up periods of ∼4 to 24 days. (A) Selected Mlxtran model code: logistic growth with the carrying capacity held fixed at $K = 1800\,\mathrm{mm}^3$ rather than estimated; $k$, growth-rate constant; $V_0$, initial tumor volume, both carrying inter-individual variability (IIV); $\omega$, standard deviation of the log-normal random effect; $b$, proportional residual error. (B) Population parameter estimates with relative standard errors (RSE) and 95% confidence intervals (CI) from SAEM (stochastic approximation expectation–maximisation) estimation. (C) Individual fits on the training set: observed (points) and individual-predicted (lines) tumor volumes. (D, E) Visual predictive checks (VPC) on the training and held-out sets: observed data against the model-predicted median and 10th/90th percentile intervals (PI, prediction interval; empirical vs. theoretical). Panels C–E share axes framed on the observations, so the prediction intervals may be cut off. (F) Observed versus individual-predicted tumor volumes on the held-out set, with the identity line and a Loess trend. Data source: Vaghi et al., PLoS Comput. Biol. 16:e1007178 (2020).

# Supplementary Figure 6: Real PD platelet cohort — $^{177}$Lu-DOTATATE (Lutathera)

A

```
[LONGITUDINAL]
input = {PLT0, ktr, slope, gamma}

PK:
depot(target=Ac)

EQUATION:
t_0 = 0
Ac_0 = 0
Prol_0 = PLT0
Tr1_0 = PLT0
Tr2_0 = PLT0
Tr3_0 = PLT0
Tr4_0 = PLT0
Tr5_0 = PLT0
PLT_0 = PLT0

; Drug concentration proxy
C = Ac

; Drug effect
Edrug = slope*C

; Feedback
feedback = (PLT0/PLT)^gamma

; Transit model with 5 compartments
ddt_Ac = -ktr*Ac
ddt_Prol = ktr*Prol*feedback*(1 - Edrug) - ktr*Prol
ddt_Tr1 = ktr*Prol - ktr*Tr1
ddt_Tr2 = ktr*Tr1 - ktr*Tr2
ddt_Tr3 = ktr*Tr2 - ktr*Tr3
ddt_Tr4 = ktr*Tr3 - ktr*Tr4
ddt_Tr5 = ktr*Tr4 - ktr*Tr5
ddt_PLT = ktr*Tr5 - ktr*PLT

OUTPUT:
output = {PLT}
```

B

| Parameter | Unit | Value | RSE (%) | 95% CI |
|---|---|---|---|---|
| *PLT0_pop* | $G.L^{-1}$ | 241 | 4.67 | [220, 264] |
| *ktr_pop* | $day^{-1}$ | 0.412 | 17 | [0.297, 0.571] |
| *slope_pop* | $MBq^{-1}$ | 3.42e-05 | 11.5 | [2.73e-05, 4.27e-05] |
| *gamma_pop* | — | 0.0268 | 27.4 | [0.0161, 0.0447] |
| *omega_PLT0* | — | 0.298 | 11.2 | [0.24, 0.371] |
| *omega_ktr* | — | 0.776 | 15.2 | [0.579, 1.04] |
| *omega_slope* | — | 0.538 | 18.7 | [0.376, 0.769] |
| *omega_gamma* | — | 1.14 | 13.8 | [0.874, 1.49] |
| *b* | — | 0.119 | 3.75 | [0.111, 0.128] |

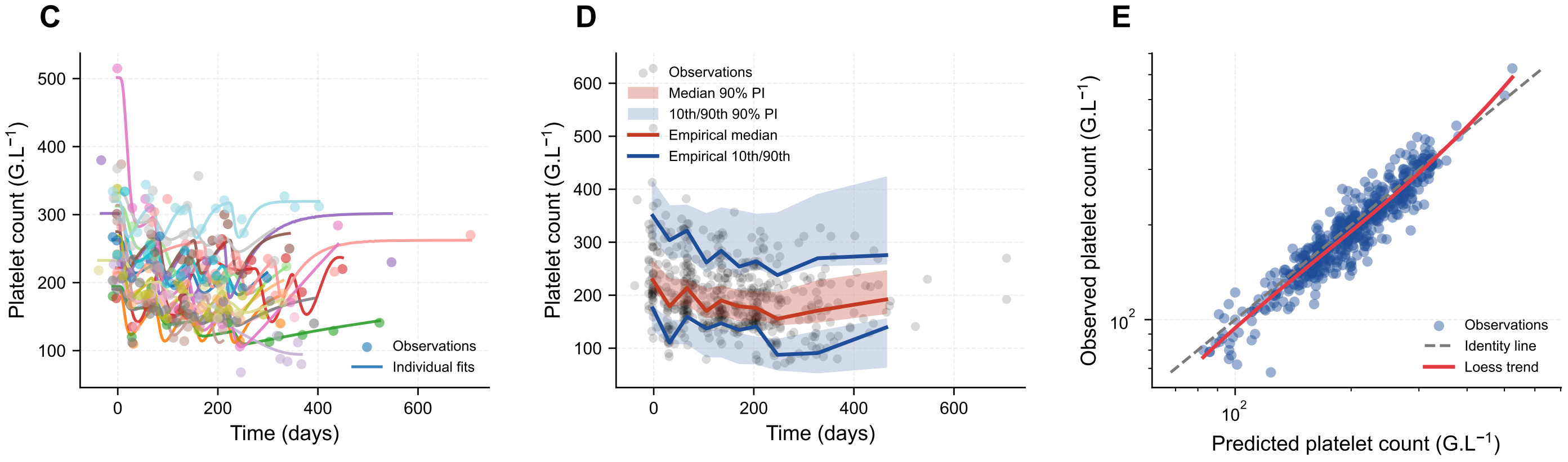


**Legend.** APOD-discovered Friberg-type myelosuppression model for the real platelet-count (pharmacodynamic, PD) training cohort ($n = 44$ patients, 519 observations) under $^{177}$Lu-DOTATATE (Lutathera) radioligand therapy, from a blinded prompt. (A) Selected Mlxtran model code: a proliferating progenitor compartment with a power-law feedback $(PLT_0/PLT)^{\gamma}$, ($PLT$, circulating platelet count; $PLT_0$, its baseline), a chain of five maturation transit compartments governed by a single transition rate $k_{tr}$, and a linear drug effect $(slope{\cdot}C)$ inhibiting proliferation. (B) Population parameter estimates with relative standard errors (RSE) and 95% confidence intervals (CI) from SAEM (stochastic approximation expectation–maximization) estimation. (C) Individual fits: observed (points) and individual-predicted (lines) platelet counts for each patient. (D) Visual predictive check: observed data against the model-predicted median and 10th/90th percentile intervals (PI, prediction interval; empirical vs. theoretical). (E) Observed versus individual-predicted platelet counts, with the identity line and a Loess trend. Platelet count in G/L ($= 10^9$/L); time in days.

# Supplementary Figure 7: Individual platelet-trajectory forecasts

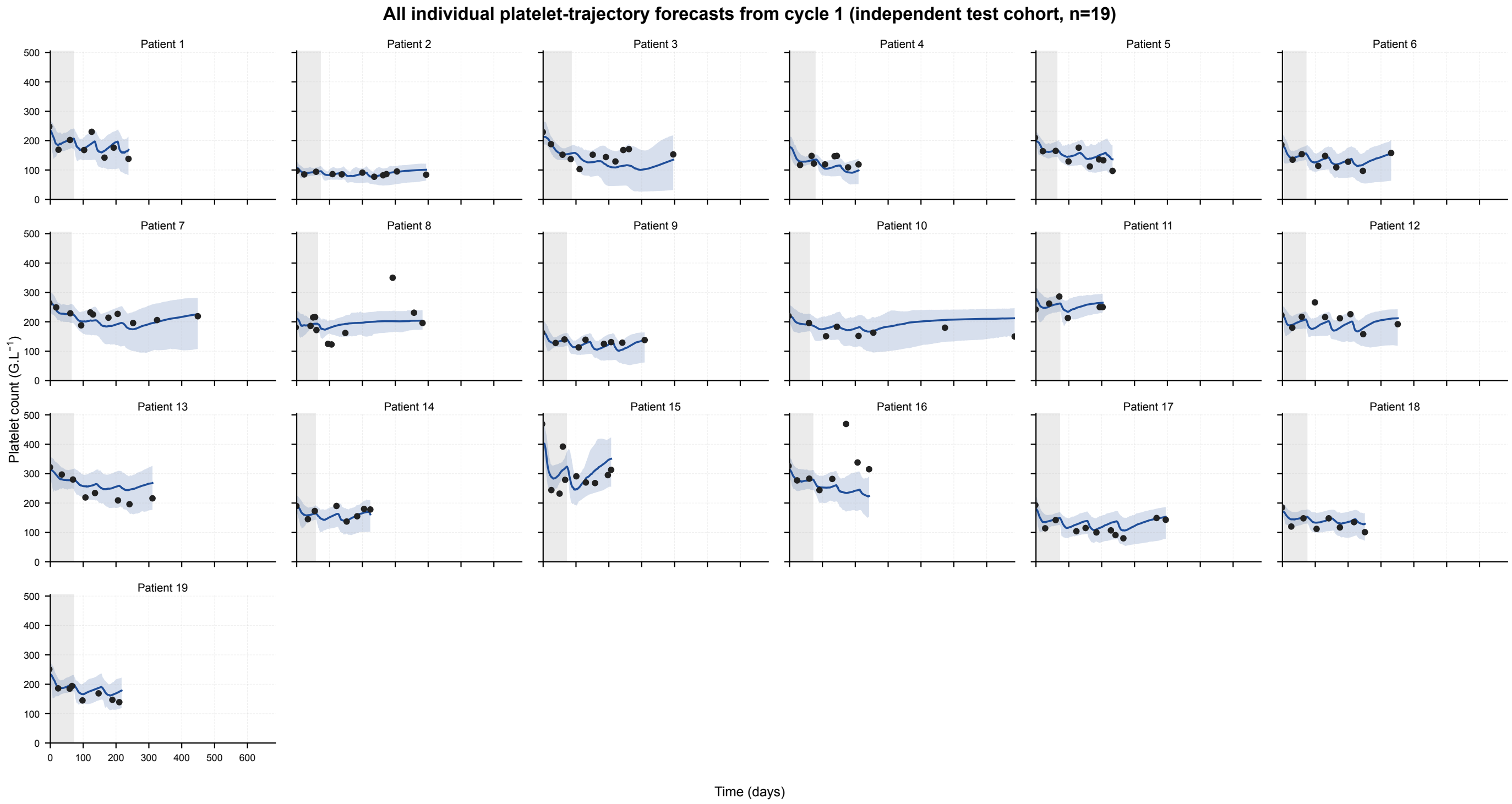


**Legend.** Individual platelet-count forecasts for all 19 patients of the independent test cohort, using the APOD-discovered Friberg-type myelosuppression model. For each patient the model is conditioned on the first treatment cycle only (grey shaded window) and used to forecast the subsequent platelet trajectory. Dots: observed platelet counts; solid line: posterior-median forecast; shaded band: 90% predictive interval (Metropolis–Hastings). Platelet count in G/L (= $10^9$/L); time in days.

## Supplementary Table 1: Structural models for the synthetic PK benchmark

| Model | Equation | Parameters (unit) | Interpretation |
|---|---|---|---|
| **1-cmt IV linear** | $\frac{dA_c}{dt} = -\frac{CL}{V} \cdot A_c$ | $CL$ (L.h$^{-1}$)<br>$V$ (L) | *Mono-exponential* |
| **2-cmt IV Michaelis-Menten** | $\frac{dA_c}{dt} = -\frac{V_m \cdot (A_c/V_1)}{K_m + A_c/V_1} - Q \cdot (\frac{A_c}{V_1} - \frac{A_p}{V_2})$<br>$\frac{dA_p}{dt} = Q \cdot (\frac{A_c}{V_1} - \frac{A_p}{V_2})$ | $V_1$ (L)<br>$V_2$ (L)<br>$Q$ (L.h$^{-1}$)<br>$V_m$ (mg.h$^{-1}$)<br>$K_m$ (mg.L$^{-1}$) | *Saturable elimination + distribution* |
| **2-cmt IV linear** | $\frac{dA_c}{dt} = -\frac{CL}{V_1} \cdot A_c - Q \cdot (\frac{A_c}{V_1} - \frac{A_p}{V_2})$<br>$\frac{dA_p}{dt} = Q \cdot (\frac{A_c}{V_1} - \frac{A_p}{V_2})$ | $CL$ (L.h$^{-1}$)<br>$V_1$ (L)<br>$V_2$ (L)<br>$Q$ (L.h$^{-1}$) | *Bi-exponential* |
| **1-cmt oral linear** | $\frac{dA_d}{dt} = -K_a \cdot A_d$<br>$\frac{dA_c}{dt} = K_a \cdot A_d - \frac{CL}{V} \cdot A_c$ | $CL$ (L.h$^{-1}$)<br>$V$ (L)<br>$K_a$ (h$^{-1}$) | *First-order absorption* |
| **1-cmt oral lag** | $\frac{dA_d}{dt} = -K_a \cdot A_d$<br>$\frac{dA_c}{dt} = K_a \cdot A_d - \frac{CL}{V} \cdot A_c$ | $CL$ (L.h$^{-1}$)<br>$V$ (L)<br>$K_a$ (h$^{-1}$)<br>$T_{lag}$ (h) | *Absorption with lag time* |
| **2-cmt oral lag** | $\frac{dA_d}{dt} = -K_a \cdot A_d$<br>$\frac{dA_c}{dt} = K_a \cdot A_d - \frac{CL}{V_2} \cdot A_c - Q \cdot (\frac{A_c}{V_2} - \frac{A_p}{V_3})$<br>$\frac{dA_p}{dt} = Q \cdot (\frac{A_c}{V_2} - \frac{A_p}{V_3})$ | $CL$ (L.h$^{-1}$)<br>$V_2$ (L)<br>$V_3$ (L)<br>$Q$ (L.h$^{-1}$)<br>$K_a$ (h$^{-1}$)<br>$T_{lag}$ (h) | *Bi-exponential + delayed absorption* |
| **2-cmt oral linear** | $\frac{dA_d}{dt} = -K_a \cdot A_d$<br>$\frac{dA_c}{dt} = K_a \cdot A_d - \frac{CL}{V_2} \cdot A_c - Q \cdot (\frac{A_c}{V_2} - \frac{A_p}{V_3})$<br>$\frac{dA_p}{dt} = Q \cdot (\frac{A_c}{V_2} - \frac{A_p}{V_3})$ | $CL$ (L.h$^{-1}$)<br>$V_2$ (L)<br>$V_3$ (L)<br>$Q$ (L.h$^{-1}$)<br>$K_a$ (h$^{-1}$) | *Bi-exponential + first-order absorption* |
| **1-cmt IV Michaelis-Menten** | $\frac{dA_c}{dt} = -\frac{V_m \cdot (A_c/V)}{K_m + A_c/V}$ | $V$ (L)<br>$V_m$ (mg.h$^{-1}$)<br>$K_m$ (mg.L$^{-1}$) | *Michaelis-Menten elimination* |
| **3-cmt IV linear** | $\frac{dA_c}{dt} = -\frac{CL}{V_1} \cdot A_c - Q_2 \cdot (\frac{A_c}{V_1} - \frac{A_{p2}}{V_2}) - Q_3 \cdot (\frac{A_c}{V_1} - \frac{A_{p3}}{V_3})$<br>$\frac{dA_{p2}}{dt} = Q_2 \cdot (\frac{A_c}{V_1} - \frac{A_{p2}}{V_2})$<br>$\frac{dA_{p3}}{dt} = Q_3 \cdot (\frac{A_c}{V_1} - \frac{A_{p3}}{V_3})$ | $CL$ (L.h$^{-1}$)<br>$V_1$ (L)<br>$V_2$ (L)<br>$V_3$ (L)<br>$Q_2$ (L.h$^{-1}$)<br>$Q_3$ (L.h$^{-1}$) | *Tri-exponential* |
| **3-cmt IV Michaelis-Menten** | $\frac{dA_c}{dt} = -\frac{V_m \cdot (A_c/V_1)}{K_m + A_c/V_1} - Q_2 \cdot (\frac{A_c}{V_1} - \frac{A_{p2}}{V_2}) - Q_3 \cdot (\frac{A_c}{V_1} - \frac{A_{p3}}{V_3})$<br>$\frac{dA_{p2}}{dt} = Q_2 \cdot (\frac{A_c}{V_1} - \frac{A_{p2}}{V_2})$<br>$\frac{dA_{p3}}{dt} = Q_3 \cdot (\frac{A_c}{V_1} - \frac{A_{p3}}{V_3})$ | $V_1$ (L)<br>$V_2$ (L)<br>$V_3$ (L)<br>$V_m$ (mg.h$^{-1}$)<br>$K_m$ (mg.L$^{-1}$)<br>$Q_2$ (L.h$^{-1}$)<br>$Q_3$ (L.h$^{-1}$) | *Saturable elimination, two peripheral compartments* |

**Legend.** Ten pharmacokinetic (PK) archetypes underlying the synthetic PK benchmark, reproduced exactly from the generative Mlxtran models used for simulation (RsSimulx). cmt: compartment (as in the model names, e.g. 1-cmt, 2-cmt, 3-cmt). $A_c$: central-compartment amount; $A_p$, $A_{p2}$, $A_{p3}$: peripheral-compartment amounts; $A_d$: absorption depot, with $dA_d/dt = -K_a A_d$ and dose delivered into $A_d$ (delayed by $T_{lag}$ where present). Central concentration $C_c = A_c/V_c$, with $V_c = V$ (1-compartment), $V_1$ (2-/3-compartment intravenous, IV), or $V_2$ (2-compartment oral). $CL$: clearance; $V$, $V_1$, $V_2$, $V_3$: volumes of distribution; $Q$, $Q_2$, $Q_3$: inter-compartmental clearances; $K_a$: absorption rate constant; $V_m$, $K_m$: Michaelis-Menten maximum elimination rate and Michaelis constant. Each parameter carries its own unit, in negative-exponent notation.

## Supplementary Table 2: Simulation design for the synthetic PK benchmark

| Model | Population parameters | IIV (ω) | n | Design (dose arms, sampling) |
|---|---|---|---|---|
| **1-cmt IV linear** | $CL = 2.25$ L.h$^{-1}$, $V = 15$ L<br>$b = 0.15$ | $\omega_{CL} = 0.25$, $\omega_V = 0.20$ | 60 | IV infusion<br>85 / 175 / 350 mg<br>8 obs, 0.5-48 h |
| **2-cmt IV Michaelis-Menten** | $V_1 = 40$ L, $V_2 = 30$ L<br>$Q = 2$ L.h$^{-1}$, $V_m = 25$ mg.h$^{-1}$<br>$K_m = 5$ mg.L$^{-1}$, $b = 0.15$ | $\omega_{V_1} = 0.20$, $\omega_{V_2} = 0.25$<br>$\omega_Q = 0.25$, $\omega_{V_m} = 0.30$<br>$\omega_{K_m} = 0.20$ | 60 | IV infusion<br>500 / 1000 / 1500 mg<br>10 obs, 0.5-72 h |
| **2-cmt IV linear** | $CL = 6$ L.h$^{-1}$, $V_1 = 15$ L<br>$V_2 = 10$ L, $Q = 5$ L.h$^{-1}$<br>$b = 0.15$ | $\omega_{CL} = 0.30$, $\omega_{V_1} = 0.20$<br>$\omega_{V_2} = 0.30$, $\omega_Q = 0.35$ | 60 | IV infusion<br>130 / 500 / 850 mg<br>8 obs, 0.25-36 h |
| **1-cmt oral linear** | $CL = 2.8$ L.h$^{-1}$, $V = 35$ L<br>$K_a = 1.5$ h$^{-1}$, $b = 0.15$ | $\omega_{CL} = 0.30$, $\omega_V = 0.20$<br>$\omega_{K_a} = 0.40$ | 60 | Oral<br>150 / 400 / 800 mg<br>8 obs, 0.25-24 h |
| **1-cmt oral lag** | $CL = 14$ L.h$^{-1}$, $V = 300$ L<br>$K_a = 1$ h$^{-1}$, $T_{lag} = 0.4$ h<br>$b = 0.15$ | $\omega_{CL} = 0.30$, $\omega_V = 0.25$<br>$\omega_{K_a} = 0.30$, $\omega_{T_{lag}} = 0.20$ | 60 | Oral<br>200 / 400 / 800 mg<br>10 obs, 0.5-48 h |
| **2-cmt oral lag** | $CL = 20$ L.h$^{-1}$, $V_2 = 50$ L<br>$V_3 = 300$ L, $Q = 30$ L.h$^{-1}$<br>$K_a = 2$ h$^{-1}$, $T_{lag} = 0.3$ h<br>$b = 0.15$ | $\omega_{CL} = 0.30$, $\omega_{V_2} = 0.15$<br>$\omega_{V_3} = 0.20$, $\omega_Q = 0.25$ | 60 | Oral<br>1 / 5 / 20 mg<br>11 obs, 0.25-72 h |
| **2-cmt oral linear** | $CL = 3.95$ L.h$^{-1}$, $V_2 = 13.1$ L<br>$V_3 = 452$ L, $Q = 7.5$ L.h$^{-1}$<br>$K_a = 0.95$ h$^{-1}$, $b = 0.15$ | $\omega_{CL} = 0.35$, $\omega_{V_2} = 0.25$<br>$\omega_{V_3} = 0.30$, $\omega_Q = 0.30$ | 60 | Oral<br>25 / 150 / 450 mg<br>10 obs, 0.25-168 h |
| **1-cmt IV Michaelis-Menten** | $V = 40$ L, $V_m = 500$ mg.h$^{-1}$<br>$K_m = 65$ mg.L$^{-1}$, $b = 0.15$ | $\omega_V = 0.25$, $\omega_{V_m} = 0.30$<br>$\omega_{K_m} = 0.30$ | 60 | IV infusion<br>500 / 1500 / 4000 mg<br>8 obs, 0.5-72 h |
| **3-cmt IV linear** | $CL = 35$ L.h$^{-1}$, $V_1 = 4$ L<br>$V_2 = 12$ L, $V_3 = 100$ L<br>$Q_2 = 40$ L.h$^{-1}$, $Q_3 = 8$ L.h$^{-1}$<br>$a = 0.003$ mg.L$^{-1}$, $b = 0.15$ | $\omega_{CL} = 0.30$, $\omega_{V_1} = 0.25$<br>$\omega_{V_2} = 0.30$, $\omega_{V_3} = 0.35$<br>$\omega_{Q_2} = 0.30$, $\omega_{Q_3} = 0.35$ | 60 | IV infusion<br>75 / 130 / 200 mg<br>9 obs, 0.05-48 h |
| **3-cmt IV Michaelis-Menten** | $V_1 = 12.8$ L, $V_2 = 60$ L<br>$V_3 = 252$ L, $V_m = 31.9$ mg.h$^{-1}$<br>$K_m = 0.453$ mg.L$^{-1}$, $Q_2 = 14$ L.h$^{-1}$<br>$Q_3 = 20.1$ L.h$^{-1}$, $b = 0.20$ | $\omega_{V_1} = 0.18$, $\omega_{V_2} = 0.28$<br>$\omega_{V_3} = 0.23$, $\omega_{V_m} = 0.16$<br>$\omega_{K_m} = 0.20$, $\omega_{Q_2} = 0.20$<br>$\omega_{Q_3} = 0.30$ | 60 | IV infusion<br>30 / 175 / 350 mg<br>11 obs, 0.5-96 h |

**Legend.** Population parameter values, inter-individual variability (IIV, expressed as $\omega$ coefficients of variation), and dosing/sampling design used to simulate each of the ten synthetic pharmacokinetic (PK) datasets (drug_01–drug_10; structural models and parameter symbols as in Supplementary Table 1; IV, intravenous). All ten datasets share the same cohort size ($n = 60$ individuals, split across three dose arms of 20). $b$ is the proportional residual error, 0.15 for every dataset except drug_10 (0.20); drug_09 additionally has an additive component $a$. IIV terms are reported only for parameters on which variability was actually applied in the generative model: no $\omega$ is listed for $K_a$ (drug_06, drug_07) or $T_{lag}$ (drug_06), which were simulated without inter-individual variability.

## Supplementary Table 3: Structural models for the synthetic TD benchmark

| Model | Equation | Parameters (unit) | Interpretation |
|---|---|---|---|
| **Exponential** | $\frac{dTS}{dt} = k_g \cdot TS$ | $TS_0$ ($mm^3$)<br>$k_g$ ($day^{-1}$) | *Unbounded exponential growth* |
| **Gompertz** | $\frac{dTS}{dt} = \alpha \cdot TS \cdot \log(\frac{K}{TS})$ | $TS_0$ ($mm^3$)<br>$\alpha$ ($day^{-1}$)<br>$K$ ($mm^3$) | *Decelerating growth* |
| **Logistic** | $\frac{dTS}{dt} = k_g \cdot TS \cdot (1 - \frac{TS}{TS_{max}})$ | $TS_0$ ($mm^3$)<br>$k_g$ ($day^{-1}$)<br>$TS_{max}$ ($mm^3$) | *Saturation* |
| **Generalized logistic** | $\frac{dTS}{dt} = k_g \cdot TS \cdot (1 - (\frac{TS}{TS_{max}})^{\gamma})$ | $TS_0$ ($mm^3$)<br>$k_g$ ($day^{-1}$)<br>$TS_{max}$ ($mm^3$)<br>$\gamma$ (–) | *Flexible saturation* |
| **Power law** | $\frac{dTS}{dt} = k_g \cdot TS^{\gamma}$ | $TS_0$ ($mm^3$)<br>$k_g$ ($(mm^3)^{1-\gamma}.day^{-1}$)<br>$\gamma$ (–) | *Sub/super-exponential* |

**Legend.** Five tumor-dynamics (TD) models used for the synthetic benchmark. $TS$: tumor size; $TS_0$: initial tumor size; $k_g$: growth-rate constant (shared notation across all models); $\alpha$, $\beta$: Gompertz deceleration parameters; $TS_{max}$: carrying capacity (Logistic, Generalized logistic); $\gamma$: shape parameter controlling the saturation profile (Generalized logistic) or sub/super-exponential growth (Power law).

# Supplementary Table 4: Simulation design for the synthetic TD benchmark

| Model | Population parameters | IIV (ω) | n | Times (days) |
|---|---|---|---|---|
| **Exponential** | $TS_0 = 15\ \text{mm}^3$<br>$k_g = 0.15\ \text{day}^{-1}$<br>$b = 0.1$ | $\omega_{TS_0} = 0.2$<br>$\omega_{k_g} = 0.1$ | 30 | 0-30<br>(every 2 days) |
| **Gompertz** | $TS_0 = 15\ \text{mm}^3$<br>$\alpha = 0.15\ \text{day}^{-1}$<br>$K = 1200\ \text{mm}^3$<br>$b = 0.1$ | $\omega_{TS_0} = 0.2$<br>$\omega_{\alpha} = 0.1$<br>$\omega_K = 0.1$ | 30 | 0-30<br>(every 2 days) |
| **Logistic** | $TS_0 = 15\ \text{mm}^3$<br>$TS_{max} = 1500\ \text{mm}^3$<br>$k_g = 0.22\ \text{day}^{-1}$<br>$b = 0.1$ | $\omega_{TS_0} = 0.2$<br>$\omega_{TS_{max}} = 0.1$<br>$\omega_{k_g} = 0.1$ | 30 | 0-30<br>(every 2 days) |
| **Generalized logistic** | $TS_0 = 15\ \text{mm}^3$<br>$TS_{max} = 1500\ \text{mm}^3$<br>$k_g = 0.3\ \text{day}^{-1}$<br>$\gamma = 0.5$<br>$b = 0.1$ | $\omega_{TS_0} = 0.2$<br>$\omega_{TS_{max}} = 0.1$<br>$\omega_{k_g} = 0.1$<br>$\omega_{\gamma} = 0.1$ | 30 | 0-30<br>(every 2 days) |
| **Power law** | $TS_0 = 15\ \text{mm}^3$<br>$k_g = 0.5\ (\text{mm}^3)^{0.25}.\text{day}^{-1}$<br>$\gamma = 0.75$<br>$b = 0.1$ | $\omega_{TS_0} = 0.2$<br>$\omega_{k_g} = 0.05$<br>$\omega_{\gamma} = 0.05$ | 30 | 0-30<br>(every 2 days) |

**Legend.** Population parameter values and inter-individual variability (IIV, $\omega$) used to simulate each of the five synthetic tumor-dynamics (TD) datasets, one per structural model in Supplementary Table 3, whose notation applies (TS, tumor size). $b$ is the proportional residual error, common to all datasets. All datasets share the same cohort size ($n = 30$ individuals) and observation schedule (tumor size sampled every 2 days over a 30-day window).

# Supplementary Table 5: Language-model backends, per dataset

| | Dataset | Claude Opus 4.5 API | | Gemini 3.1 Pro API | | GPT-5.6-sol API | | DeepSeek-V4 Pro API | | GPT-OSS 20B local, 20B | | Qwen3 8B local, 8B | | Mistral Small 3.2 local, 24B | |
|---|---|---|---|---|---|---|---|---|---|---|---|---|---|---|---|
| | | N tested (failed %) | Time (min) | N tested (failed %) | Time (min) | N tested (failed %) | Time (min) | N tested (failed %) | Time (min) | N tested (failed %) | Time (min) | N tested (failed %) | Time (min) | N tested (failed %) | Time (min) |
| PK | 1-cmt IV linear | 7 (0%) | 2.9 | 6 (0%) | 7.8 | 5 (0%) | 4.9 | 7 (30%) | 4.7 | 25 (88%) | 10.5 | 7 (36%) | 9.9 | 22 (67%) | 22.2 |
| | 2-cmt IV MM | 13 (0%) | 5.7 | 12 (0%) | 8.3 | 3 (0%) | 1.7 | 6 (25%) | 9.2 | 16 (81%) | 14.0 | 5 (47%) | 7.6 | 18 (73%) | 14.5 |
| | 2-cmt IV linear | 9 (0%) | 3.9 | 9 (0%) | 6.1 | 4 (0%) | 2.4 | 5 (22%) | 4.6 | 18 (79%) | 16.7 | 6 (14%) | 6.7 | 16 (72%) | 11.8 |
| | 1-cmt oral linear | 10 (3%) | 4.1 | 9 (0%) | 5.7 | 6 (0%) | 4.0 | 3 (24%) | 3.2 | 22 (84%) | 15.8 | 4 (36%) | 9.8 | 9 (100%) | -- |
| | 1-cmt oral + lag | 9 (3%) | 4.1 | 13 (12%) | 9.9 | 21 (0%) | 28.8 | 11 (44%) | 11.5 | 23 (86%) | 21.6 | 5 (57%) | 8.6 | 14 (93%) | 13.0 |
| | 2-cmt oral + lag | 15 (1%) | 8.9 | 10 (0%) | 12.1 | 16 (4%) | 28.2 | 14 (32%) | 11.9 | 25 (86%) | 19.5 | 10 (40%) | 18.5 | 14 (86%) | 16.5 |
| | 2-cmt oral linear | 9 (0%) | 6.4 | 11 (2%) | 11.2 | 6 (0%) | 3.1 | 5 (20%) | 3.8 | 16 (77%) | 11.0 | 4 (86%) | 12.0 | 9 (89%) | 6.0 |
| | 1-cmt IV MM | 18 (0%) | 10.1 | 12 (2%) | 9.3 | 20 (0%) | 26.2 | 14 (28%) | 15.6 | 14 (85%) | 11.3 | 8 (43%) | 9.8 | 12 (80%) | 8.9 |
| | 3-cmt IV linear | 9 (0%) | 5.1 | 6 (0%) | 6.3 | 5 (6%) | 6.4 | 8 (14%) | 11.0 | 24 (85%) | 22.7 | 8 (31%) | 9.8 | 20 (68%) | 14.8 |
| | 3-cmt IV MM | 12 (0%) | 9.3 | 16 (0%) | 27.1 | 4 (0%) | 1.9 | 8 (29%) | 6.1 | 21 (83%) | 20.2 | 10 (44%) | 13.2 | 13 (70%) | 12.7 |
| TD | Exponential | 5 (0%) | 1.3 | 11 (0%) | 4.9 | 4 (0%) | 1.3 | 6 (28%) | 4.2 | 6 (19%) | 2.6 | 4 (26%) | 3.4 | 20 (47%) | 10.8 |
| | Logistic | 10 (1%) | 2.3 | 11 (10%) | 3.8 | 5 (0%) | 1.3 | 6 (19%) | 3.3 | 3 (0%) | 1.0 | 3 (5%) | 2.3 | 29 (54%) | 10.8 |
| | Gompertz | 10 (0%) | 2.5 | 7 (3%) | 2.8 | 4 (0%) | 1.1 | 4 (22%) | 2.3 | 3 (17%) | 3.4 | 5 (20%) | 7.2 | 25 (49%) | 9.0 |
| | Generalized logistic | 15 (4%) | 3.9 | 16 (10%) | 8.2 | 13 (2%) | 6.7 | 6 (12%) | 5.6 | 10 (31%) | 4.2 | 4 (15%) | 2.9 | 33 (48%) | 15.6 |
| | Power law | 15 (1%) | 3.6 | 12 (5%) | 5.2 | 8 (0%) | 3.8 | 9 (31%) | 3.8 | 10 (36%) | 4.2 | 9 (18%) | 5.3 | 34 (56%) | 14.9 |
| | *Across all datasets* | 10 (1%) | 4.1 | 10 (4%) | 7.0 | 5 (1%) | 3.1 | 7 (28%) | 5.1 | 17 (75%) | 11.3 | 5 (32%) | 6.5 | 21 (58%) | 11.5 |

**Legend.** The same discovery loop driven by seven language-model backends on the ten pharmacokinetic (PK) and five tumor-dynamics (TD) datasets of Table 2 (cmt, compartment; IV, intravenous; MM, Michaelis–Menten), one column block per backend; Table 3 gives the structures recovered in these cells. N tested: median number of candidate structures evaluated per run, with the percentage of candidates that could not be fitted in parentheses, almost always invalid Mlxtran syntax. Time: median end-to-end run time in minutes over the runs of the cell. "Across all datasets": the same two statistics taken over every run of the backend rather than per dataset – median candidates per run, and the share of all those candidates that failed – both restricted, as the cells are, to the runs that attempted at least one candidate; runs that died before writing a candidate are charged to the backend in Table 3, whose denominator is the number of repeats planned. The median application programming interface (API) cost per run is in Table 3. "–": no completed run.

# Supplementary Table 6: Parameter estimates for the real cohorts

**A. Docetaxel (PK)**

| Parameter | Unit | Value | RSE (%) | 95% CI |
|---|---|---|---|---|
| $V_1$ | L | 7.38 | 5.83 | [6.59, 8.27] |
| $k_{10}$ | $h^{-1}$ | 5.22 | 4.7 | [4.76, 5.72] |
| $k_{12}$ | $h^{-1}$ | 0.959 | 0.462 | [0.95, 0.967] |
| $k_{21}$ | $h^{-1}$ | 1.07 | 0.48 | [1.06, 1.08] |
| $k_{13}$ | $h^{-1}$ | 1.1 | 5.61 | [0.989, 1.23] |
| $k_{31}$ | $h^{-1}$ | 6.88e-02 | 7.1 | [5.99e-02, 7.91e-02] |
| $\omega_{V1}$ | — | 0.217 | 19.6 | [0.149, 0.315] |
| $\omega_{k10}$ | — | 0.184 | 17.8 | [0.13, 0.258] |
| $\omega_{k13}$ | — | 0.198 | 22.4 | [0.13, 0.303] |
| $b$ | — | 0.204 | 4.32 | [0.188, 0.223] |

**B. Breast cancer xenografts (TD)**

| Parameter | Unit | Value | RSE (%) | 95% CI |
|---|---|---|---|---|
| $\lambda$ | $day^{-1}$ | 0.116 | 1.94 | [0.111, 0.12] |
| $K$ | $mm^3$ | 1.71e+03 | 5.1 | [1.54e+03, 1.88e+03] |
| $\gamma$ | — | 0.599 | 1.05 | [0.587, 0.611] |
| $TS_0$ | $mm^3$ | 3.46 | 17.5 | [2.47, 4.83] |
| $\omega_K$ | — | 0.242 | 15.7 | [0.179, 0.328] |
| $\omega_{TS0}$ | — | 0.9 | 12.2 | [0.71, 1.14] |
| $b$ | — | 0.179 | 4.11 | [0.165, 0.194] |

**C. Subcutaneous polymer-prodrug (PK-TD)**

| Parameter | Unit | Value | RSE (%) | 95% CI |
|---|---|---|---|---|
| $k_g$ | $h^{-1}$ | 3.52e-03 | 3.63 | [3.28e-03, 3.78e-03] |
| $K_{max}$ | $mm^3$ | 2.95e+03 | 9.44 | [2.45e+03, 3.54e+03] |
| $k_d$ | $mL.ng^{-1}.h^{-1}$ | 6.88e-02 | 9.8 | [5.69e-02, 8.33e-02] |
| $C_{th}$ | $ng.mL^{-1}$ | 1.08e-02 | 20.2 | [7.33e-03, 1.58e-02] |
| $k_{e0}$ | $h^{-1}$ | 1.13e-03 | 9.68 | [9.35e-04, 1.36e-03] |
| $TS_0$ | $mm^3$ | 39 | 2.92 | [36.8, 41.3] |
| $\omega_{kg}$ | — | 0.283 | 10.1 | [0.232, 0.344] |
| $\omega_{Kmax}$ | — | 0.345 | 43.4 | [0.161, 0.737] |
| $\omega_{kd}$ | — | 0.457 | 15.8 | [0.337, 0.619] |
| $\omega_{Cth}$ | — | 0.796 | 36.1 | [0.415, 1.53] |
| $\omega_{ke0}$ | — | 0.536 | 12.6 | [0.42, 0.685] |
| $\omega_{TS0}$ | — | 0.163 | 14 | [0.125, 0.214] |
| $b$ | — | 0.301 | 2.36 | [0.287, 0.315] |

**D. Platelets, radioligand therapy (PD)**

| Parameter | Unit | Value | RSE (%) | 95% CI |
|---|---|---|---|---|
| $PLT_0$ | $G.L^{-1}$ | 2.41e+02 | 4.67 | [2.20e+02, 2.64e+02] |
| $k_{tr}$ | $day^{-1}$ | 0.412 | 17 | [0.297, 0.571] |
| $slope$ | $MBq^{-1}$ | 3.42e-05 | 11.5 | [2.73e-05, 4.27e-05] |
| $\gamma$ | — | 2.68e-02 | 27.4 | [1.61e-02, 4.47e-02] |
| $\omega_{PLT0}$ | — | 0.298 | 11.2 | [0.24, 0.371] |
| $\omega_{ktr}$ | — | 0.776 | 15.2 | [0.579, 1.04] |
| $\omega_{slope}$ | — | 0.538 | 18.7 | [0.376, 0.769] |
| $\omega_\gamma$ | — | 1.14 | 13.8 | [0.874, 1.49] |
| $b$ | — | 0.119 | 3.75 | [0.111, 0.128] |

**Legend.** Population parameter estimates of the APOD-discovered model for each real cohort of Figures 2 and 3, with relative standard errors (RSE) and 95% confidence intervals (CI) from the stochastic approximation of the Fisher information matrix. One sub-table per cohort: (A) docetaxel plasma pharmacokinetics (PK), three-compartment intravenous disposition; (B) untreated breast-cancer xenograft tumor dynamics (TD), hyper-Gompertz growth law; (C) the coupled pharmacokinetic–tumor-dynamic (PK-TD) polymer-prodrug cohort, logistic growth with a concentration-threshold kill term driven by an effect compartment; (D) platelet dynamics (pharmacodynamics, PD) under radioligand therapy. Sub-tables A–C are the three rows of Figure 2 and D is the cohort of Figure 3; the fitted model is the one the figure shows, resolved from the same run, so for the three cohorts split into training and held-out sets these are the training-set estimates. $\omega_x$: standard deviation of the log-normal random effect on $x$; $b$: proportional residual error. Parameter symbols: $V_1$, central volume, $k_{10}$, elimination rate constant, $k_{12}$, $k_{21}$, $k_{13}$, $k_{31}$, inter-compartmental rate constants (A); $\lambda$, growth-rate constant, $K$, carrying capacity, $\gamma$, shape exponent, $TS_0$, initial tumor size (B); $k_g$, growth-rate constant, $K_{max}$, carrying capacity, $k_d$, kill-rate constant, $C_{th}$, effect-site concentration threshold of the kill term, $k_{e0}$, effect-compartment rate constant, $TS_0$, initial tumor size (C); $PLT_0$, baseline platelet count, $k_{tr}$, transit rate constant, $slope$, linear drug effect, $\gamma$, feedback exponent (D). Random effects and $b$ are dimensionless, as is the shape exponent $\gamma$ of the hyper-Gompertz. Time is in days for the tumor-volume cohorts and in hours for the PK–TD cohort. A cohort absent from the results tree is listed at the foot of the table rather than silently omitted.

## Supplementary Table 7: Patient characteristics and platelet outcome of the radioligand-therapy cohorts

| Characteristic | Training cohort (n = 44) | Test cohort (n = 19) |
|---|---|---|
| **Patient characteristics** | | |
| Age (years), mean (s.d.) | 64 (12) | 70 (9) |
| Female sex, n (%) | 22 (50) | 8 (42) |
| Ki-67 (%), mean (s.d.) | 7.9 (6.4) | 8.4 (5.2) |
| Metastatic disease, n (%) | 44 (100) | 19 (100) |
| **Primary tumour site, n (%)** | | |
| Small intestine | 27 (61) | 15 (79) |
| Lung | 5 (11) | 3 (16) |
| Pancreas | 2 (5) | 0 (0) |
| Other site | 6 (14) | 0 (0) |
| Unknown primary | 4 (9) | 1 (5) |
| **Treatment** | | |
| Activity per cycle (MBq) | 7,400 | 7,400 |
| Cycles received, median (range) | 4 (1–4) | 4 (2–4) |
| **Platelet monitoring** | | |
| Platelet counts, n | 519 | 172 |
| Counts per patient, median [IQR] | 12 [9–14] | 9 [8–11] |
| Baseline count ($G.L^{-1}$), median [IQR] | 224 [202–294] | 220 [186–250] |
| Nadir ($G.L^{-1}$), median [IQR] | 139 [114–189] | 137 [102–173] |
| Nadir ($G.L^{-1}$), range | 68–307 | 77–244 |
| **CTCAE grade at nadir, n (%)** | | |
| Grade 0 | 17 (39) | 7 (37) |
| Grade 1 | 25 (57) | 12 (63) |
| Grade 2 | 2 (5) | 0 (0) |
| Grade 3 or 4 | 0 (0) | 0 (0) |

**Legend.** Characteristics of the two cohorts of patients with metastatic neuroendocrine tumours treated with [$^{177}$Lu]Lu-DOTATATE at Institut Paoli-Calmettes: the training cohort, on which the myelosuppression model was estimated, and the independent test cohort, on which first-cycle forecasts were evaluated. Every patient received a fixed activity of 7,400 MBq per cycle, for up to four cycles eight weeks apart. The nadir is the lowest platelet count recorded for a patient over the whole treatment course. Grades follow the Common Terminology Criteria for Adverse Events (CTCAE) v5.0 for *platelet count decreased*, taking 150 G/L as the lower limit of normal: grade 1, 75–150 G/L; grade 2, 50–75 G/L; grade 3, 25–50 G/L; grade 4, below 25 G/L; no patient in either cohort reached grade 3. Ki-67 was unavailable for 1 patient in the training cohort and 4 in the test cohort. IQR, interquartile range; s.d., standard deviation.



\end{document}